\documentstyle[11pt,aaspp4]{article}
\slugcomment{Accepted by AJ February 18 2004, to appear in June 2004 issue}

\lefthead{Knapp et al.}
\righthead{}

\begin{document}
 
\title{Near-Infrared Photometry and Spectroscopy of L and T Dwarfs: the Effects
of Temperature, Clouds, and Gravity}

\author{G.~R.\ Knapp\altaffilmark{\ref{Princeton}},
S.~K.\ Leggett\altaffilmark{\ref{UKIRT}},
X. Fan\altaffilmark{\ref{Steward}},
M.~S. Marley \altaffilmark{\ref{NASA}},
T.~R.\ Geballe\altaffilmark{\ref{Gemini}},
D. A.\ Golimowski\altaffilmark{\ref{JHU}},
D. Finkbeiner\altaffilmark{\ref{Princeton}},
J. E.\ Gunn\altaffilmark{\ref{Princeton}},
J. Hennawi\altaffilmark{\ref{Princeton}},
\v{Z}. Ivezi\'{c}\altaffilmark{\ref{Princeton}},
R. H.\ Lupton\altaffilmark{\ref{Princeton}},
D. J. Schlegel\altaffilmark{\ref{Princeton}},
M. A.\ Strauss\altaffilmark{\ref{Princeton}},
Z. I.\ Tsvetanov\altaffilmark{\ref{JHU}},
K. Chiu\altaffilmark{\ref{JHU}},
E.A. Hoversten\altaffilmark{\ref{JHU}},
K. Glazebrook\altaffilmark{\ref{JHU}},
W. Zheng\altaffilmark{\ref{JHU}},
M. Hendrickson\altaffilmark{\ref{JHU}},
C. C. Williams\altaffilmark{\ref{JHU}},
A. Uomoto\altaffilmark{\ref{JHU},\ref{OCIW}},
F.J. Vrba\altaffilmark{\ref{USNOFS}},
A.A. Henden\altaffilmark{\ref{USNOFS},\ref{USRA}},
C.B. Luginbuhl\altaffilmark{\ref{USNOFS}},
H.H. Guetter\altaffilmark{\ref{USNOFS}},
J.A. Munn\altaffilmark{\ref{USNOFS}},
B. Canzian\altaffilmark{\ref{USNOFS}},
Donald P. Schneider\altaffilmark{\ref{PennState}},\\
J. Brinkmann\altaffilmark{\ref{APO}}
%Suzanne L.\ Hawley\altaffilmark{\ref{UW}}
%Istv\'{a}n Csabai\altaffilmark{\ref{JHU},\ref{Eotvos}},
%Todd J.\ Henry\altaffilmark{\ref{JHU},\ref{GSU}},\\
%Robert Hindsley\altaffilmark{\ref{NRL}},
%Jeffrey R. Pier\altaffilmark{\ref{USNOFS}},
%J.\ Allyn Smith\altaffilmark{\ref{Wyoming}},
%D.~G.\ York\altaffilmark{\ref{Chicago}}
\newcounter{address}
\setcounter{address}{1}
\altaffiltext{\theaddress}{Princeton University Observatory, Princeton, 
NJ 08544
\label{Princeton}}
\addtocounter{address}{1}
\altaffiltext{\theaddress}{United Kingdom Infrared Telescope, Joint Astronomy
Centre, 660 North A'ohoku Place, Hilo, Hawaii 96720 
\label{UKIRT}}
\addtocounter{address}{1}
\altaffiltext{\theaddress}{ 
Steward Observatory, 933 N. Cherry Ave. Tucson, AZ 85721
\label{Steward}}
\addtocounter{address}{1}
\altaffiltext{\theaddress}{NASA Ames Research Center,
Mail Stop 245-3, Moffett Field, CA 94035
\label{NASA}}
\addtocounter{address}{1}
\altaffiltext{\theaddress}{Gemini Observatory, 670 North A'ohoku Place,
Hilo, HI 96720
\label{Gemini}}
\addtocounter{address}{1}
\altaffiltext{\theaddress}{
Department of Physics and Astronomy, The Johns Hopkins University,
   3400 North Charles Street, Baltimore, MD 21218 
\label{JHU}}
\addtocounter{address}{1}
\altaffiltext{\theaddress}{
Carnegie Observatories, 813 Santa Barbara Street, Pasadena, California 91101
\label{OCIW}}
\addtocounter{address}{1}
\altaffiltext{\theaddress}{US Naval Observatory, Flagstaff Station, 
P.~O. Box 1149, Flagstaff, AZ 86002-1149
\label{USNOFS}}
\addtocounter{address}{1}
\altaffiltext{\theaddress}{Universities Space Research Association, 1101
17th St., NW, Washington, DC 20036
\label{USRA}}
\addtocounter{address}{1}
\altaffiltext{\theaddress}{Department of Astronomy and Astrophysics,
The Pennsylvania State University,
University Park, PA 16802
\label{PennState}}
\addtocounter{address}{1}
\altaffiltext{\theaddress}{Apache Point Observatory, 2001 Apache Point Road, 
P.O.~Box 59, Sunspot, NM 88349
\label{APO}}
%\addtocounter{address}{1}
%\altaffiltext{\theaddress}{Department of Astronomy, University of Washington,
%Box 351580, Seattle, WA 98195
%\label{UW}}
%\addtocounter{address}{1}
%\altaffiltext{\theaddress}{Department of Physics and Complex Systems, 
%E\"{o}tv\"{o}s
%University, P\'{a}zm\'{a}ny P\'{e}ter s\'{e}t\'{a}ny 1/A, Budapest, H-1117, 
%Hungary
%\label{Eotvos}}
%\addtocounter{address}{1}
%\altaffiltext{\theaddress}{Department of Physics and Astronomy, Georgia State 
%University,
%Atlanta, GA  30303
%\label{GSU}}
%\addtocounter{address}{1}
%\altaffiltext{\theaddress}{Remote Sensing Division,
%United States Naval Research Laboratory, 
%Washington, DC 20375
%\label{NRL}}
%\addtocounter{address}{1}
%\altaffiltext{\theaddress}{Department of Physics and Astronomy,
%University of Wyoming, 
%P.O.\ Box 3905, Laramie, WY 82071
%\label{Wyoming}}
%\addtocounter{address}{1}
%\addtocounter{address}{1}
%\altaffiltext{\theaddress}{University of Chicago, Astronomy \& Astrophysics
%Center, 5640 S. Ellis Ave., Chicago, IL 60637
%\label{Chicago}}
}

\begin{abstract}

We present new $JHK$ photometry on the MKO-NIR system and $JHK$ spectroscopy 
for a large sample of L and T dwarfs. Photometry has been obtained for 71 
dwarfs and spectroscopy for 56. The sample comprises newly identified very red
objects from the Sloan Digital Sky Survey (SDSS) and known dwarfs from the
SDSS and the Two Micron All Sky Survey (2MASS). Spectral classification
has been carried out using four previously defined indices (from Geballe
et al. 2002, G02) that measure the strengths of the near infrared water and 
methane bands.  We identify 9 new L8--9.5 dwarfs and 14 new T dwarfs from 
SDSS, including the latest yet found by SDSS, the T7 dwarf SDSS 
J175805.46$+$463311.9. We classify 2MASS J04151954$-$0935066 as T9, the 
latest and coolest dwarf found to date.

We combine the new results with our previously published data to
produce a sample of 59 L dwarfs and 42 T dwarfs with imaging data on a
single photometric system and with uniform spectroscopic classification.  
We compare the near-infrared colors and absolute magnitudes of brown dwarfs 
near the L--T transition with predictions made by models of the distribution 
and evolution of photospheric condensates.  There is some scatter in the G02 
spectral indices for L dwarfs, suggesting that these indices are probing 
different levels of the atmosphere and are affected by the location of the 
condensate cloud layer.  The near-infrared colors of the L dwarfs also show 
scatter within a given spectral type, which is likely due to variations in 
the altitudes, spatial distributions and thicknesses of the clouds. We have 
identified a small group of late L dwarfs that are relatively blue for their
spectral type and that have enhanced FeH, H$_2$O and K~I absorption, 
possibly due to an unusually small amount of condensates.

The scatter seen in the $H-K$ color for late T dwarfs can
be reproduced by models with a range in surface gravity.  The variation is 
probably due to the effect on the $K$-band flux of pressure-induced $\rm
H_2$ opacity. The correlation of $H-K$ color with gravity is supported by
the observed strengths of the $J$-band K~I doublet. Gravity is closely
related to mass for field T dwarfs with ages $>10^8$~yrs and the gravities
implied by the $H-K$ colors indicate that the T dwarfs in our sample have
masses in the range 15 -- 75 $\rm M_{Jupiter}$. One of the SDSS dwarfs,
SDSS J111010.01$+$011613.1, is possibly a very low mass object, with log $g$
$\sim$ 4.2 -- 4.5 and mass $\sim$ 10 -- 15 $\rm M_{Jupiter}$.

\end{abstract}
 
\section{Introduction}

Since the discovery of dwarfs of spectral type later than M as companions 
to nearby stars (Becklin \& Zuckerman 1988; Nakajima et al. 1995) major 
observational and theoretical progress has been made, thanks to sensitive 
new wide-area surveys at optical (0.4--1.0~$\mu$m) and infrared 
(1.0--2.5~$\mu$m) wavelengths and models of atmospheres of dwarfs with 
effective temperatures between those of the coolest stars and the giant 
planets (see the reviews by Chabrier \& Baraffe 2000; Burrows et al. 2001). 
Two new spectral classes have been identified later than type M: the L 
dwarfs, characterized by the disappearance of gas-phase TiO and VO, and 
the T dwarfs, characterized by methane band absorption in the $H$ and $K$ 
spectral regions (Mart\'{\i}n et al. 1997, 1999b; Kirkpatrick et al. 1999, 2000; 
Strauss et al. 1999; Leggett et al. 2000, 2002b; Geballe et al. 2002, 
hereafter G02; Burgasser et al. 2002a; Hawley et al. 2002).  Most L dwarfs 
and all T dwarfs are brown dwarfs.  These objects are of interest because 
they occupy the mass range between that of stars and giant planets; because 
many of them are likely to have the intrinsic properties of giant planets, 
which at present cannot be directly observed; and because they allow the 
investigation of the initial mass function to substellar masses.

Field L and T dwarfs have been discovered in large numbers in recent sky
surveys: the Deep Near Infrared Survey (DENIS, Epchtein 1997); the Two
Micron All Sky Survey (2MASS, Skrutskie et al. 1997; Beichman et al.
1998), and the optical Sloan Digital Sky Survey SDSS (York et al. 2000).  
Including objects described in the present paper, there are now about 280
L dwarfs and 58 T dwarf systems published (e.g. Delfosse
et al. 1997, 1999;  Kirkpatrick et al. 1999, 2000;  Burgasser et al. 2002a, 2003e; 
G02). This large sample has been used to establish a complete spectral sequence 
from L0 to T9 (G02; Burgasser et al. 2002a; McLean et al. 2003;  present paper).

Unlike stars, brown dwarfs lack a sustained source of thermonuclear
energy, and hence cool continuously, passing through the L and T stages, with
their initial spectral types depending on their masses. Their observational
properties are thus a function not only of mass and metallicity, but also
of age. Not all dwarfs of a given spectral type or effective temperature
are identical; they have different gravities and different colors, the
latter likely due to differing amounts of particulate matter in the
atmosphere.  For example, in mid to late L dwarfs there is a large scatter in
the $JHK$ colors and apparently no one-to-one correspondence between
effective temperature and spectral type (Leggett et al. 2002a; Golimowski
et al. 2004).  Such considerations drive searches for and measurements of
additional dwarfs, in order to more fully characterize their atmospheres.

2MASS and SDSS have been highly complementary  in the
discovery of L and T dwarfs.  Most of the flux from late-type dwarfs is
emitted longward of 1 $\mu$m, and the $J-H$ and $H-K$ colors of M and L
dwarfs become redder with decreasing effective temperature, allowing the
identification by 2MASS of large numbers of L dwarfs (Kirkpatrick et al.
1999, 2000).  However, in the transition from L to T, $\rm CH_4$
absorption appears in the $H$ and $K$ regions (and $\rm H_2$ absorption
predominantly at $K$), strengthening with later spectral type and causing
the T dwarfs to become increasingly blue in their $JHK$
colors.  The $JHK$ colors of early T dwarfs are similar to those of the
common K and M stars, making their identification in 2MASS very difficult.
However in the SDSS filters the dwarfs simply become redder and thus
the early T dwarfs have been found primarily in SDSS imaging (Leggett et
al. 2000).  Only three objects with spectral types between T0 and T3.5 have 
been identified from sources other than the SDSS -- Mart\'{\i}n et al. (2001)
in an optical and near-infrared imaging survey of the $\sigma$ Orionis cluster
tentatively classified one member as T0, Liu et al. (2002) found a distant field
T3--T4 dwarf in a deep $Iz$ survey, and McCaughrean et al. (2003)
found that $\epsilon$ Indi B, discovered in a high proper motion optical survey, 
is a binary consisting of a T1 and T6 pair (see also Scholz et al. 2003, 
Smith et al. 2003 and Volk et al. 2003).  To date, all T
dwarfs later than T7 have been found in the 2MASS database (Burgasser et
al. 2002a) but we expect such objects to be found in the SDSS imaging data
as sky coverage is increased.

In this paper we present near-infrared photometry and spectra of new and
previously reported L and T dwarfs (including 14 new T dwarfs) and compare
the colors and spectra with predictions from state of the art model
atmospheres with and without clouds.  The new objects observed are
described in the next section, and the new $JHK$ photometric and
spectroscopic observations are described in \S 3, where we also derive
spectral types.  \S 4 presents colors, spectral types and absolute
magnitudes for the entire body of near-infrared data on L and T dwarfs
which we have accumulated to date.  In \S 5, we compare these data to
model atmospheres. The conclusions are given in \S 6.

\section{The Observed Sample}

The general goals of our observational efforts are to identify samples of
late L and T dwarfs which are at least representative and ideally
complete, and to measure their spectral and photometric characteristics.  
To this end, we have observed new candidate very cool dwarfs selected from
the photometric observations of the SDSS.  We have also re-observed some
previously published SDSS L and T dwarfs for which our observations are
incomplete or suspect, and have added some 2MASS dwarfs where they
complement our sample.  Previous observations of the 2MASS objects are
discussed by Kirkpatrick et al. (1999, 2000), Reid et al. (2001),
Burgasser et al. (2002a,b; 2003b,c) and Dahn et al. (2002).

Table 1 lists the names and SDSS $i,z$ photometry of 51 confirmed SDSS
dwarfs, most of them previously unpublished.  A small number of
these dwarfs are also detected in the SDSS $r$ band, and these magnitudes  
are presented later in the paper.  As noted in Table 1, a few of these
objects are described by Hawley et al. (2002), who present optical spectral 
types of a large sample of M, L and T dwarfs
observed by SDSS.  One object, 2MASS J090837.97$+$503208.0, was identified
as an L dwarf by Cruz et al. (2003) from the 2MASS database while this paper 
was in preparation.  Finding charts from the SDSS $z$-band imaging are 
presented in Figure 1 for all 51 objects in Table 1 as charts have not been
previously published for any of the objects.

Following the IAU convention, the SDSS names are based on the J2000
coordinates at the epoch of the initial observation and will be abbreviated 
in the text when individual objects are discussed; thus SDSS 
J003259.36$+$141036.6 will be called by the shortened name SDSS J0032$+$1410. 
A similar convention is used for 2MASS objects, whose full coordinate names 
are given later (in Table 9). Note that these faint stars and brown dwarfs 
are nearby and likely to have significant proper motions, so the 
names do not necessarily provide accurate coordinates for later epochs.
A future paper will discuss the measured SDSS positions and proper motions, 
since many of the objects were measured at more than one epoch. 

The SDSS $i$ and $z$ magnitudes are $\rm AB_{\nu}$ magnitudes, for which the 
zeropoint in all bands is 3631 Jy (Oke \& Gunn 1983; Fukugita et al. 1996).  
The $ZJHK$ magnitudes used at the United Kingdom Infrared Telescope
(UKIRT) and discussed in this paper are on the $m$(Vega)$=$
0.0 system.  The SDSS
magnitudes are modified to be asinh magnitudes, identical to logarithmic
magnitudes for high signal-to-noise ratio measurements ($>5\sigma$) and
linear with flux below this (Lupton, Gunn \& Szalay 1999).  For the
observations discussed herein, zero flux density corresponds to $r =$
25.1, $i =$ 24.4 and $z =$ 22.8.

The SDSS camera (Gunn et al. 1998) scans the sky and produces
near-simultaneous CCD images in five filters covering the optical bands
$u,g,r,i$ and $z$ (centered at 3551 \AA{}, 4686 \AA{}, 6166 \AA{}, 7480
\AA{} and 8932 \AA{}.) The imaging data are reduced through a set of
automated software pipelines.  The photometric pipeline Photo (Lupton et
al. 2002) corrects the data, finds and measures
objects, and applies photometric and astrometric calibrations. The
photometric calibration is provided via a network of standard stars (Hogg
et al. 2001; Smith et al. 2002), and the astrometric calibration via
matches to standard astrometric catalogues (Pier et al. 2003). The
photometry is accurate to about 2\% in $g,r$ and $i$ and to about 3\% in
$u$ and $z$ for objects brighter than about 20 and 19 respectively, while
the astrometric accuracy is better than $0.1''$ (r.m.s.) in each
coordinate. The result is a catalogue of objects with magnitudes in five
bands, positions, and shape parameters (e.g. Abazajian et al. 2003 and
references therein).

All L and T dwarfs are undetected in SDSS $u$ and $g$, and all save the
brightest are undetected in $r$. Almost all late L and T dwarfs have $i-z
>$ 2; thus candidate field brown dwarfs are selected from the SDSS
photometry to be very red. Given the extremely red colors and low
luminosities of these dwarfs, they are often detected only in the $z$
band.  Objects this red are rare; at the SDSS magnitude limits
they are late L and T dwarfs, quasars at redshift greater than 5.7 (Fan et
al. 2001, 2003) or very rare unusual broad absorption line quasars (see
Hall et al. 2002), and their surface density is smaller than about one per 50
square degrees.  As a result, $z$-band only detections are overwhelmed by
data artifacts, in particular ``cosmic rays'' in the $z$ detectors.  The
winnowing of ``objects'' to find those that are real is an exhaustive
process, consisting of careful analysis of the image to reject cosmic rays
(which usually have imprints smaller than the point
spread function), re-observation with another telescope at $z$ band
(usually the ARC 3.5 m telescope at Apache Point Observatory), and comparison 
with $J$-band observations from 2MASS or elsewhere.  $J$-band photometry 
also allows a first judgment as to whether an object is a brown dwarf or a 
high redshift quasar.  These procedures are described in detail by Fan et
al. (2001, 2003). The total area searched to date for candidate very red dwarfs 
in SDSS, including objects in our previous papers and in Table 1, is 2870
square degrees, so the surface density of T dwarfs found by SDSS is
approximately 1 per 100 square degrees.  The regions of the sky searched
are shown by Fan et al. (2003).

Some of the objects in the final sample presented in \S 4 have been found 
to be close binaries, usually from $HST$
imaging.  Such objects will be designated by ``AB'' attached to the names, since
the photometric and spectroscopic observations measure the total flux of
both members.  These known binaries are DENIS-P~J0205$-$1159AB (Koerner et
al. 1999; Leggett et al. 2001; Bouy et al. 2003); DENIS-P~J1228$-$1547AB 
(Koerner et al. 1999; Mart\'{\i}n et al. 1999a; Bouy et al. 2003); 
2MASS J0746$+$2000AB and 2MASS J0850$+$1057AB (Reid et al. 2001; Bouy et al. 2003);  
2MASS J1225$-$2739AB and 2MASS J1534$-$2952AB (Burgasser et al.
2003d); and 2MASS J1553$+$1532AB (A. Burgasser, private communication,
2003). Not all of the objects in the sample have been imaged at high
angular resolution, however, and there are likely to be more binaries
among them.

\section{New Observational Data}

\subsection{Near-Infrared Photometry}

Table 2 gives new $Z$ photometry and Table 3 new $JHK$ photometry. The
central wavelengths of the filter passbands are 0.95, 1.25, 1.64 and
2.2~$\mu$m; more details of the filters and calibration are given by
Leggett et al. (2002a). All data were obtained on UKIRT. All $Z$-band 
data were obtained with UKIRT's Fast-Track Imager (UFTI, Roche et al. 
2003) on the dates shown in Table 2.  The $JHK$ data were taken with the 
Mauna Kea consortium filter set (MKO-NIR) on the dates and with the cameras 
as listed in Table 3.  Three cameras were used for the $JHK$ observations 
--- the InfraRed Camera (IRCAM, McLean et al. 1986), UFTI, and the UKIRT 
Imager-Spectrometer (UIST, Ramsay-Howat et al. 2000). UFTI contains a 
HAWAII 1024$\times$1024 HgCdTe detector and has a plate scale of 
0$\farcs$091 pixel$^{-1}$. IRCAM contains an SBRC 256$\times$256 InSb 
detector and has a plate scale of 0$\farcs$081 pixel$^{-1}$. UIST contains 
an ALADDIN 1024$\times$1024 InSb detector and has a choice of plate scales, 
either 0$\farcs$061 pixel$^{-1}$ or 0$\farcs$120 pixel$^{-1}$.  The 
0$\farcs$120 pixel$^{-1}$ plate scale was used with UIST for these $JHK$ 
observations.  Readout of the full IRCAM array was employed, but 
512$\times$512 subarrays were used with UFTI and UIST to reduce overheads 
(i.e. to increase efficiency).   The fields of view were thus 20$\farcs$7, 
46$\farcs$6 and 61$\farcs$4 for IRCAM, UFTI, and UIST respectively.

Individual exposure times were usually 250 seconds at $Z$ and 60 seconds
at each of $JHK$. Observations were made in a three or five position
dither pattern at $Z$, and with five or nine dither positions at $JHK$.  
The $JHK$ data were calibrated using the UKIRT Faint Standards of Hawarden
et al. (2001) translated onto the MKO-NIR system using as yet unpublished
observations carried out at UKIRT as part of an observatory project to
provide calibrators in the MKO-NIR system. The $Z$ data were calibrated
using unpublished UKIRT observations.  These calibration data are currently 
available via the UKIRT web 
pages\footnote{http://www.jach.hawaii.edu/JACpublic/UKIRT/astronomy/calib/fs\_izjhklm.dat}.

We have investigated the effects of the different optical elements, their
coatings and the detector anti-reflection coatings on the $JHK$
photometric systems of the three cameras.  Synthesizing $JHK$ for L and T
dwarfs using flux-calibrated spectra (see G02) shows that the differences
at $J$ are about 0.009 magnitudes for L to early T types, and 0.013
magnitudes for late T dwarfs. At $H$ the differences are around 0.001
magnitudes for objects of both L and T spectral type. At $K$, the
difference is 0.001 magnitudes for objects of type L, 0.003 magnitudes for
objects of type early T, and 0.010 magnitudes for late T dwarfs.  In all
cases this is significantly less than the measurement error, so that the
data from the three cameras are effectively on the same photometric
system, defined by the MKO-NIR filter set.  Transformations between this
filter set and other widely-used $JHK$ filter sets (e.g. the 2MASS system)
are described by Stephens \& Leggett (2004).

\subsection{Spectroscopy and Spectral Types}

Table 4 lists the instrument configurations for the new spectroscopic
observations. All spectra were obtained at UKIRT using either the Cooled
Grating Spectrometer (CGS4, Wright et al. 1993) or UIST.  CGS4 has a SBRC
256$\times$256 InSb detector with 0$\farcs$6 pixels. In UIST's
spectroscopy mode the ALADDIN array has 0$\farcs$12 pixels. Individual
exposure times were typically 120 seconds for the CGS4-Z and CGS4-J
settings, 60 seconds for CGS4-H and CGS4-K and 120 seconds for UIST-HK.  
The targets were nodded 7--12 arcseconds along the slit.  A- or early-to-mid 
F-type bright stars were used as calibrators to remove the effects of the
terrestrial atmosphere, with H I recombination lines in their spectra
removed artificially prior to ratioing.  Both instruments have 
lamps that provide accurate flatfielding and wavelength calibration.

A log of the measured spectra is given in Table 5. We concentrated on  
obtaining spectra in the $H$- and $K$-bands, because indices in these 
wavelength regions can be used for a wide range of types (G02). Table 6 
gives the derived spectral indices and the mean implied type 
on the G02 scheme; the individual classifications are rounded off to the 
nearest 0.5 of a subclass but the mean type is derived from the unrounded 
values.  Errors are given for those dwarfs with multiple indices which show 
a scatter larger than the estimated classification uncertainty of 0.5
subclasses.  The reader is referred to G02 for examples of the spectral
sequences and line identifications.  Spectra and photometry from this
and our previous papers are available on request or from our L and T dwarf
web page\footnote{http://www.jach.hawaii.edu/$\sim$skl/LTdata.html}.
Note that unlike G02 we do not incorporate red  ``PC3'' and ``Color-d''
spectral indices for the  current sample.  These indices can be used for 
classifying dwarfs of spectral type L6 and earlier (G02); however, in this 
paper we present near-infrared data only.  A future paper will examine red 
spectra where available for the sample, and investigate the 
wavelength-dependent effects of cloud condensation (see discussion later 
in \S 5.2).

Table 6 contains 14 new T dwarfs, including one SDSS object optically 
classified as L8 by Hawley et al. (2002).  The total number of T dwarfs 
presently known is 58, four of which are close binaries.  The distribution of
new SDSS T dwarfs is: two T0, three T1--T1.5, two T2, two T4.5--T5, two T5.5, 
two T6, and one T7.  The last of these, SDSS J1758$+$4633, is the latest-type 
dwarf found to date in the SDSS.  Nine L8--L9.5 dwarfs have been identified
from new infrared spectra.  Together with the seven new early-T dwarfs, they
significantly increase the number of known dwarfs in the L--T transition region.
Finally, the near-infrared photometry for SDSS J1649$+$3842 and the photometry 
and spectroscopy for SDSS J0747$+$2937 show that they are M dwarfs.  The CH$_4$-K 
index for the latter (see Table 6) is at the limit of the G02 scheme and is very 
uncertain.  These M dwarfs are not discussed further.  We have not obtained 
infrared spectra for six of the 51 objects listed in Table 1.

The spectrum of 2MASS J0415$-$0935 has significantly deeper $\rm H_2O$ and
$\rm CH_4$ bands than previously known T8 dwarfs and we classify it as T9;
it is the latest spectral type dwarf presently known.  Figure 2 shows the
$H$- and $K$-band spectra of SDSS J1758$+$4833 (T7) and 2MASS J0415$-$0935 (T9),
in addition to our previously-published spectra of the T6 dwarf
SDSS J1624$+$0029 (Strauss et al. 1999) and the T8 dwarf Gl 570D (Geballe et
al. 2001).  The steady increase in the depths of the $\rm H_2O$ and $\rm
CH_4$ bands from T6 to T9 can be seen.  Figure 2 suggests that there is
room for one more T type which would have essentially zero flux at
1.45~$\mu$m, 1.7~$\mu$m, and 2.25~$\mu$m.  Provisional indices for the end 
of the T sequence are given in Table 7.

According to the models of Burrows, Sudarsky \& Lunine (2003), $\rm NH_3$ is
expected to be detectable in the $H$ and $K$ spectral regions (at $\sim$ 
1.5 $\mu$m, 1.95 $\mu$m and 2.95 $\mu$m) for $\rm T_{eff} \leq$ 600~K, and its
presence may mark the transition to the spectral type after T --- although new
non-equilibrium chemistry models suggest that the abundance of $\rm NH_3$ 
may be reduced (Saumon et al. 2003).  As we discuss in our companion paper
(Golimowski et al. 2004), the effective temperature of 2MASS J0415-0935 is 
$\sim$700~K, too warm for $\rm NH_3$ absorption. There is no sign of 
$\rm NH_3$ absorption in the spectrum (Figure 2), or in any other $HK$ spectra
we have obtained to date.

\section{The Final Sample}

We have compiled a large sample of L and T dwarfs for further study 
by combining the new data presented in \S 3 with our previously published
work (Strauss et al. 1999; Fan et al. 2000; Tsvetanov et al. 2000;
Leggett et al. 2000, 2001, 2002a,b;  Geballe et al. 2001, 2002). This final 
sample consists of 63 spectroscopically confirmed L dwarfs 
(59 of which have infrared spectra), six other possible L dwarfs measured
photometrically only, and 42 spectroscopically confirmed T dwarfs.

%\subsection{Absolute Magnitudes}

The distances to 45 of these L and T dwarfs are known by virtue of recent
parallax measurements or because they are companions to nearby stars with
accurately-measured parallaxes, either from the ground (van Altena et al.
1995) or from $Hipparcos$ (ESA 1997, Perryman et al. 1997).  Since the
discovery of isolated L and T dwarfs, much effort has been devoted to the
measurement of accurate parallaxes, both at optical wavelengths (Tinney et
al. 1995, Dahn et al. 2002) and recently at near-infrared wavelengths
(Tinney et al. 2003; Vrba et al. 2004).  Table 8 presents available
parallaxes of L and T dwarfs for which we have obtained UKIRT data,
together with the derived $K$-band luminosities on the MKO system.  Some
of the parallaxes are weighted mean values from more than one source, as noted
in Table 8.  The errors for $M_K$ given in the table are the combined
errors in the parallax and in the photometry.

%\subsection{Final Sample}

Table 9 summarizes our final sample of L and T dwarfs.  Column 1 gives full
coordinate names for the SDSS and 2MASS dwarfs; these are listed as footnotes 
for dwarfs discovered in other work (e.g. for Kelu--1, Ruiz et al. 1997).   
Column 2 lists spectral types from the following sources: this paper (using 
spectra described in Table 6 or presented by Burgasser et al. 2002a); G02; 
Kirkpatrick et al. (1999, 2000); and Burgasser et al. (2002a). The uncertainty 
in the assigned type is given if there are multiple infrared spectral indices 
which deviate by more than the estimated classification uncertainty of 0.5 
subclasses.  The spectral type is also flagged if the infrared and optical 
types (Kirkpatrick et al. 1999, 2000; Hawley et al. 2002; Cruz et al. 2003) 
differ by more than 1.0 subclass.  Discrepant indices, either between the 
optical and infrared or even within the infrared range, are a sign that the 
spectra are sampling very different regions of the atmosphere, as we discuss 
later in \S 5.2 and \S 5.7.4. Column 3 lists $M_J$, derived from our
$J$ measurements and the parallaxes in Table 8.  The next two columns list
the SDSS $i-z$ and $z$.  These values are given only  for those objects 
for which $\sigma_i$ or $\sigma_z <$ 0.2 magnitudes.  They are  based on the 
most recent SDSS reductions (Photo $5\_4\_25$, July 2003) and may differ slightly
from previously published values.  SDSS photometry for non-SDSS dwarfs, where 
available, are included in Table 9.  The remaining columns list $z-J$,  $Z-J$,
$J$, $J-K$, $J-H$ and $H-K$.  Some near-infrared magnitudes and colors are 
synthesized from the spectra, as noted in the table.  A small number of dwarfs 
have detectable SDSS $r$-band fluxes; those dwarfs for which $\sigma_r < 0.3$ mag 
are given in Table 10.  Note again that the SDSS $r$, $i$ and $z$ measurements 
are on the AB system, while the other magnitudes are on the Vega$=$0 system.
$Z$ is on the UKIRT UFTI photometric system while $J$, $H$ and $K$ are on
the MKO-NIR system.

For most of the objects in the sample, we have only single measurements of $JHK$.
However, we have obtained repeat photometry for a few objects with unusual
colors.  For each of the following dwarfs two measurements were obtained
that agree to within the observational uncertainties, and the results are
simply averaged in Table 9: 2MASS J0036$+$1821 (observed on 20001205 and
20021207); SDSS J0107$+$0041 (19991017, 20010124); SDSS J0830$+$4828 (20001119,
20001206); SDSS J0931$+$0327 (20020217, 20030104); SDSS J1104$+$5548 (20020109, 
20031204) and SDSS J1331$-$0116 (20020109, 20020620). We have discarded data 
obtained on 20000314 (Leggett et al. 2002a) for two objects.  The $J$ 
measurements for SDSS J1314$-$0008 appear to be spuriously bright; 
we have combined the $H$ and $K$ data from that night with the $JHK$ 
data presented in Table 3 to produce the colors given in Table 9.  All of the
$JHK$ data taken on that same night for SDSS J1326$-$0038 appear to be
discrepant and we have averaged data from 20010124 and 20030129 in Table
9.  Finally, data for two of the redder mid L dwarfs (2MASS J0028$+$1501 and
SDSS J2249$+$0044) suggest that they may be variable at the 5\%--10\% level,
which is not unexpected, given published detections of variability of L
dwarfs and the possibility that non-uniform clouds exist in their
atmospheres (e.g. Gelino et al. 2002; Enoch et al. 2003, and the
discussion in \S 5.7).  Including data taken with IRCAM on
20001119 that were discarded by Leggett et al. (2002a), four measurements
exist for 2MASS J0028$+$1501 and three for SDSS J2249$+$0044.  The former object
shows variations at $JHK$ of about 0.05 mag, while the latter object
varies by about 0.1 mag at each of $JHK$.  Weighted means are given for
these two objects in Table 9 and the individual datasets are listed in the
footnotes.

\section{Discussion}

\subsection{Characteristics of L and T Dwarf Atmospheres}

As effective temperatures of dwarfs cool to those of the late M dwarfs and 
below, two chemical changes occur in their photospheres that strongly
impact their emergent spectral energy distributions.  The first to occur,
for late M dwarfs, is the appearance of corundum ($\rm Al_2O_3)$ grains
within the photosphere (Jones \& Tsuji 1997) and the formation of condensate 
clouds.  At the even lower effective temperatures of the L dwarfs iron and 
silicate are the most important condensates\footnote {These condensates are
frequently termed ``dust'', but this can be misleading since in many
dwarfs the iron will be in the liquid phase (Lodders 1999). Hence we
generally prefer the terms ``condensate'' to refer to grains or drops of
the condensed phase, and ``cloud'' as the region in the atmosphere
within which the condensed species are found.}.  The effect of the clouds
is to weaken or veil the molecular absorption bands and to redden the
$JHK$ colors of L dwarfs (see e.g. Ackerman \& Marley 2001, Allard et al.
2001, Marley et al. 2002, Tsuji \& Nakajima 2003).  The extent of these
effects depends upon the number, size, and vertical distribution of the
condensates; for spectral modeling these parameters must either be
computed from a model or somehow specified.  Ackerman \& Marley (2001)
developed a one-dimensional model of mixing and sedimentation for this
purpose.  In their model upward vertical mixing of gas and condensate
replaces condensates that fall through the cloud base, while far above the
cloud base sedimentation efficiently cleanses the atmosphere of
condensates. Tsuji \& Nakajima (2003)  model the cloud by specifying the
temperature range within which the condensates are found.  They describe their
limits to be the points at which the atmosphere is cool enough for
condensation but hot enough that the condensates are small enough to
remain suspended in the atmosphere and are less prone to sedimentation.  
The temperature domain in which the cloud is formed depends on the details
of the model used, but is around $\sim$1500--1700~K (Ackerman \& Marley
2001) or $\sim$1800--2000~K (Tsuji \& Nakajima 2003). In the T dwarfs the
cloud layer lies near the base or below the wavelength-dependent
photosphere and plays a smaller role in determining the observed flux
distribution.

The second and later chemical change that occurs in these high-pressure,
low-temperature atmospheres is the formation of additional molecular
species in the photosphere, most importantly CH$_4$.  CO and H$_2$O are
abundant in M dwarf atmospheres, but by mid-L the abundance of CH$_4$
becomes significant at the expense of CO (Noll et al. 2000). At moderate
spectral resolution CH$_4$ absorption is not seen in the near-infrared
until temperatures drop to those of the late L dwarfs, at which point
$K$-band CH$_4$ features are detectable, and at T0 (by definition, G02) 
CH$_4$ absorptions are seen at both $H$ and $K$.  The increasing CH$_4$
absorption largely accounts for the increasingly blue $JHK$ colors of the
T dwarfs with later spectral type, more than compensating for the
reddening due to the decreasing effective temperature.  For dwarfs of type
T5 and later, pressure-induced H$_2$ becomes a significant
opacity source.  This opacity depresses the flux in the $K$ band, and to
a lesser extent the $H$ band, and also contributes to the blue
near-infrared colors.  For a useful summary of the important molecular
species and the wavelength ranges in which they are observed see Figure
15 of Burrows et al. (2001).

\subsection{Clouds, Molecules and Classification Schemes}

Spectral classification schemes for L and T dwarfs have been developed
using both the red and the near-infrared spectral regions.  In the late
1990s Kirkpatrick et al. (1999) and Mart\'{\i}n et al. (1999b) developed
schemes using the strengths of various absorption features and
pseudo-continuum slopes seen in optical spectra to classify the L dwarfs.
A few years later, Burgasser et al. (2002a) and G02 presented schemes using
the strengths of the near-infrared molecular absorption bands  to
classify the T dwarfs, and in the case of G02, L dwarfs also. While the
Burgasser et al. (2002a) and G02 schemes for T dwarfs give results in very
close agreement, the G02 scheme for L dwarfs can give results that differ
by as much as 2.5 subclasses from those given by the Kirkpatrick et al.
(1999) scheme, suggesting that there are significant differences in the
optical and infrared classification of L dwarfs. Dwarfs in our sample 
whose optical and infrared spectral classes differ by more than one 
subclass are identified in Table 9.  Some of the scatter is due to the 
small differences in the infrared indices from one subtype to the next, 
combined with measurement errors.  However, the differences in optical 
and infrared spectral types are not entirely random. Where there are 
differences, G02 generally assign earlier spectral types to those objects 
that are redder than average in $J-K$, and later types to those which are 
bluer (Stephens 2001, Figure 7.5).

Models of L and T dwarf atmospheres give some insight into the variations
seen among the spectral indices.  Generally speaking, the spectra of L and
T dwarfs are less sensitive to the effects of cloud decks in the
0.7--1.0~$\mu$m spectral region than at wavelengths longer than 1$\mu$m.  
This is because,
for effective temperatures corresponding to the earliest L types, optical
depth unity in the far-red is reached below the cloud deck, but as the
cloud is still fairly optically thin it does not substantially influence
the red spectrum.  At lower effective temperatures, the clouds place a
``floor'' on the region from which the emergent flux arises, but because
of the large opacity in the far-red due to initially refractory diatomics
and water and later to K~I and Na~I resonance line absorption, most of the
outgoing red flux arises from above the cloud decks.  Thus for dwarfs with
effective temperature cooler than about 1800~K (types $\sim$L3 and later,
Leggett et al. 2002a,b; Golimowski et al. 2004) slight changes in the cloud
deck optical depth have little effect on the emergent red spectrum.

In the near-infrared (particularly the $Z$ and the $J$ bands), the windows
between the molecular bands of water and other opacity sources allow flux
to emerge from very deep in the atmosphere.  In these regions the opacity
floor imposed by the clouds substantially alters the depth to which one
can see into the atmosphere (see Figure 7 of Ackerman \& Marley 2001 and
Figure 4 of Marley et al. 2002). Thus for dwarfs with $\rm T_{eff}$ in the
range from about 1800 to 1500~K (roughly L3 to L7), slight changes in the
cloud profile substantially alter the near-infrared spectrum.

This proposed atmospheric structure implies that spectral typing schemes
for mid to late L dwarfs that rely on far-red spectra (e.g.
Kirkpatrick et al. 1999) tend to be less sensitive to the vertical
distribution of condensates in the atmosphere than schemes that rely upon
near-infrared spectra or spectral indices (G02). Stephens (2001, 2003)
considered the boundaries of the regions employed in the G02 spectral
typing system and found that the flux in the G02 bandpasses usually
originates from within the cloud decks.  For the earliest L dwarfs, cloud
opacity is not significant, but for the mid L dwarfs, the G02 1.5 $\mu$m
water index can be a more sensitive indicator of {\it cloud optical depth}
than of {\it effective temperature}. At the same time, the 2.2 $\mu$m
methane index is more sensitive to $\rm T_{eff}$ than to cloud properties,
since this spectral region is more opaque and optical depth unity is
reached higher in the atmosphere. However, the classification system of
G02 relies heavily on the 1.5 $\mu$m index, as it is the only infrared
index that covers the entire L spectral type range. This index undergoes a
much larger change through the L sequence than does the other useful
infrared index, CH$_4$--K.  Note that G02 do not claim that defining the
spectral type is equivalent to measuring the effective temperature;
they use a simple classification scheme based on spectral
appearance in which the effects of gravity and clouds are not
separated from those of temperature.

This larger sample of dwarfs also suggests some inconsistency in the
infrared classification of late L dwarfs.  While the G02 scheme provides
excellent internal consistency for T dwarfs, the H$_2$O 1.5 $\mu$m index 
tends to give a later spectral type than does the CH$_4$ 2.2 $\mu$m index
for dwarfs in the range L5--L9.5.  This tendency was not apparent in the 
smaller G02 sample; the results for the present larger sample suggest that 
an adjustment of the flux ratio definitions as a function of spectral type 
for H$_2$O 1.5 $\mu$m and CH$_4$ 2.2 $\mu$m in the L5--L9.5 range could give
better internal consistency.

\subsection{Observed Color as a Function of Spectral Type}

Figure 3 plots, for the final sample presented in Table 9, various colors
against spectral type (determined by the G02 scheme apart from four L
dwarfs classified optically, see Table 9);  typical error bars are shown.  
Only those dwarfs with types determined from their spectra are shown.
Clouds strongly affect both colors and spectral types of mid L dwarfs
classified from the near-infrared indices, as discussed in \S 5.1 and \S
5.2.  This effect can be seen in Figure 3, where the spread in the $Z$
through $K$ colors is greatest from L3 to L7.5, just when the clouds in the
detectable atmosphere are expected to be most optically thick.  The
overall conclusion is that color cannot be used as an (infrared) spectral
type indicator for L dwarfs. For T dwarfs, $z-J$ and $J-H$ appear to be
reasonable indicators of type.  Given improved sensitivity, $i-z$ may also 
be a useful T-type indicator.  Note, however, that $i-z$ is expected to turn
blueward at $\rm T_{eff}~\lesssim$~600~K as the Na, K, and other alkalis
condense into solids and the opacity of the Na and K lines falls (Burrows
et al. 2002; Marley et al. 2002). The late T dwarfs show significant
scatter in their $H-K$ and $J-K$ colors. This can be understood in terms
of the onset of pressure-induced H$_2$ absorption, which is very
gravity sensitive (Borysow, Jorgensen \& Zheng, 1997).  
As described in \S 5.6, we can interpret the observed spread in $H-K$ as 
a range in surface gravity for the field T dwarf population.

The $Z-J$ colors of the mid T dwarfs also show considerable scatter
(Figure 3).  This is an intriguing result, because the $Z$ and $J$ bands
are the most transparent windows into these atmospheres, with the $Z$ band
being the clearer of the two.  As such, these bands are particularly
sensitive to any deep variations in opacity between atmospheres, such as
might be related to the upper reaches of any remaining deep silicate cloud
(see Figure 7 of Ackerman \& Marley 2001).  The variation may thus be due
to differences in the process(es) responsible for the removal of
condensates at the L to T transition. Since the $Z$ band is sensitive to
the far wings of the optical Na-D and K I resonance lines (Burrows \&
Volobuyev 2003) these variations might arise from differences in the
removal of gaseous Na and K, due to gravity or metallicity effects. More
detailed modeling of the L to T dwarf transition and the removal of
atmospheric condensates is required to account for these observations.

\subsection{Families of Models for Comparison with the Data}

% I put this in its own section so that a reader of the paper can quickly find it

In the remainder of this paper we compare observed colors with two
varieties of models from Marley et al. (2002).  In the first type of
model, condensate opacity is ignored, although condensation chemistry is
accounted for in chemical equilibrium and molecular opacities. We term
these the ``cloud-free'' models.  In the second type the effects of
condensate opacity are computed using the Ackerman \& Marley (2001) cloud
model. In these ``cloudy'' models the efficiency of condensate
sedimentation is parameterized by $f_{\rm sed}$\footnote{
Ackerman \& Marley (2001) employed the parameter ``$f_{\rm rain}$"
to describe the efficiency of condensate sedimentation in a brown dwarf
atmosphere.  Strictly speaking, ``rain'' is falling water, and this term
has now been replaced by $f_{\rm sed}$.}.  When the sedimentation
efficiency is high (large $f_{\rm sed}$) both the optical depth and vertical 
extent of the cloud are small.  In the extreme case of no condensate
sedimentation $f_{\rm sed}=0$.

\subsection{$J-H$ and $H-K$ Colors of L Dwarfs: Unusually Red and Unusually 
Blue L Dwarfs}

Figure 4 shows $J-H$ plotted against $H-K$ for the L dwarfs in the sample,
where spectral subclass ranges are indicated by different symbols.
Overlaid are cloudy model sequences with $\log g=5$ and $f_{\rm sed}$
values of 3 and 5.  The $f_{\rm sed}=3$ models match the data well,
although a shift in modeled $J-H$ color of about 0.15 mag would encompass
many more of the data points. The discrepancy is likely attributable to
the modeled TiO bands at $J$ being too deep, causing the $J$-band
magnitudes to be too faint and the $J-H$ model color to be too red.
Whether this is a shortcoming in the chemical equilibrium calculation or
the molecular opacities themselves is as yet unclear.   Log~$g=$4 models
make the $H-K$ colors bluer, in better agreement with the data, but this
is an unlikely gravity for these field dwarfs. (Burrows et al. 1997 show
that if $\rm T_{eff}$=1500--2200~K and age=1--5 Gyr then log $g\geq$5.0.)
 
The detailed distribution of most of the data points in this color-color
space is challenging to interpret as both the models and the observations
show that $J-H$ and $H-K$ first become redder and then bluer with falling
$T_{\rm eff}$.  Thus the colors double back on themselves.  Despite the
scattered distribution, some extreme objects stand out. The L7.5 dwarf
2MASS J2244$+$2043 and the L5.5 dwarf SDSS J0107$+$0041 are quite red in
both colors.  This may imply that their condensate cloud decks are more
optically thick than average, which could arise from either less efficient
sedimentation ($f_{\rm sed}\sim 2$) or higher metallicity.

Our sample also includes four late-type L dwarfs --- SDSS J0805$+$4812, 
SDSS J0931$+$0327, SDSS J1104$+$5548 and SDSS J1331$-$0116 --- that are 
unusually blue for their spectral types.  As shown in Figure 3, they
are bluer than average at $J-H$ by about 0.2~mag and at $H-K$ by about 
0.1~mag.  Applying the shift to the models described above of 
about $-0.15$ in $J-H$, Figure 4 suggests that these dwarfs are better 
described by the $f_{\rm sed}=5$ models, i.e. the sedimentation
efficiency is high and the cloud optical depth is small.
The spectral indices for all four of these objects show a large range --- 
the $H$-band indices imply a late type of  L9 to T1 while the 
$K$-band index gives an earlier type of L5.5--7.5 (the $J$-band index
only implies a type earlier than T0). Our spectra show that they have 
enhanced FeH, K~I and $\rm H_2O$ absorption (although the spectrum
of SDSS J1104$+$5548 is noisy).  Figure 5 shows the $J$-band spectra for 
SDSS J0805$+$4812, SDSS J0931$+$0327 and SDSS J1331$-$0116 bracketed by 
more typical L5.5 and L9 dwarfs.  Gorlova et al. (2003) and
McLean et al. (2003) show that the equivalent widths of the $J$-band
FeH and K~I features peak at spectral types around L3 and then become
smaller as the Fe condenses into grains and K is lost to KCl. 
The strengths of these features in SDSS J0805$+$4812, SDSS J0931$+$0327
and SDSS J1331$-$0116 are similar to those of the early L types, while their
H$_2$O bands are more typical of the latest L dwarfs, supporting the 
interpretation that we are looking deep into unusually condensate-free 
atmospheres.  The possibility of low metallicity should also be considered.
Burgasser et al.  (2003a) identified a late L dwarf (2MASS J05325346$+$8246465) 
whose extremely blue near-infrared colors are similar to those of the mid T
types.  This high velocity dwarf appears to be an extremely metal-poor
halo subdwarf with strong FeH features as well as H$_2$ absorption which 
depresses the  $H$ and $K$ band fluxes.  Cruz et al. (2003) identify two
early L dwarfs (2MASS J1300425$+$191235 and 2MASS J172139$+$334415) that are bluer 
than average at $J-K$ by about 0.2~mag.  As condensate clouds are optically 
thin for early L dwarfs, and these dwarfs have significant proper motion, 
they may also be part of a low-metallicity population.

\subsection{$J-H$ and $H-K$ Colors of T Dwarfs: Gravity and Mass 
Determinations}

The $H-K$ and $J-K$ colors of T5--T9 dwarfs are scattered (Figure 3), even 
though the $H$- and $K$-band indices of G02 yield consistent classifications.
Figure 6 shows $J-H$ against $H-K$ for the T dwarfs in the sample with 
sequences from the cloud-free and cloudy $f_{\rm sed}=5$ models by 
Marley et al. (2002) overlaid.  The synthetic $H-K$ colors for the late T 
dwarfs, and the range in color over a plausible range of gravities of 
log~$g=4.5$--5.5, reproduce the observed colors extremely well.

Gl 570 D provides a further test for the model/data correspondence shown
in Figure 6.  Geballe et al. (2001) fit models to the absolute luminosity
of Gl 570 D and used age constraints to find $\rm T_{eff}=784$--824 K and
a surface gravity in the range 5.00--5.27. This gravity range is
consistent with that implied by Figure 6.  The effective temperature
implied by Figure 6, however, is high by about $150\,\rm K$.  Although
Geballe et al. (2001) found that their best fitting models generally
reproduced the $JHK$ spectrum of Gl 570D quite well, there were notable
discrepancies.  In particular the notorious inadequacy of the $H$-band
methane opacity database and the tendency of all clear-atmosphere models
to overestimate the water-band depths limit the fidelity of the fit.  
These deficiencies both result in the best-fitting temperature contours in
Figure 6 being somewhat too warm.

Further, the trends shown in Figure 6 may break down for lower
temperatures. One interesting challenge is the T9 dwarf 2MASS J0415$-$0935,
the latest and coolest T dwarf currently known, with $\rm
T_{eff}\approx$700~K (Golimowski et al. 2004, Vrba et al. 2004). Figure 6,
8 and 9 show  that instead of being bluer in $J-H$, 
$H-K$ and $J-K$ than Gl 570D, it is {\it redder}, by 0.2 magnitudes, in $J-K$.
While models by Marley et al. (2002) and Burrows et al. (2003) predict that,
indeed, the coolest dwarfs will become redder in $J-K$ with falling 
$\rm T_{eff}$ and the onset of water cloud formation, this happens only for
models with $\rm T_{eff}~\lesssim$~500~K  unless the brown dwarf is older 
than 7 Gyr and more massive than 40 $\rm M_{Jupiter}$. The condensation of the 
alkalis into their solid chloride forms may also lead to redder colors at 
these kind of temperatures (Lodders 1999, Marley 2000, Burrows et al. 2003).

Despite these discrepancies, the overall trends seen in Figure 6 can be 
understood in the context of our presently limited understanding of brown 
dwarf atmospheres.  At the effective temperatures of late T dwarfs, the 
$K$-band flux is very sensitive to gravity.  This is because the opacity of 
pressure--induced $\rm H_2$ absorption is proportional to the square of the 
local gas number density.  Higher gravity objects of a given $\rm T_{eff}$ tend
to be cooler at a given pressure than lower gravity dwarfs, and thus have 
denser, more opaque, atmospheres.  Hence high gravity objects tend to be
dimmer at $K$ and bluer in $H-K$ than comparable lower gravity
objects. 2MASS J0937$+$2931 shows a particularly depressed $K$-band flux,
standing out in Figure 6, presumably due to strong $\rm H_2$ opacity
(Burgasser et al. 2002a). Structural models imply an upper limit
to $\log g$ of 5.5 for brown dwarfs (e.g. Burrows 1997) and therefore Figure 
6 suggests that 2MASS J0937$+$2931 may be both a high gravity {\it and} a low
metallicity dwarf, as also suggested by Burgasser et al. (2003b).  While
$\rm H_2$ opacity is also enhanced by decreasing metallicity (e.g. 
Saumon et al. 1994, Borysow et al. 1997), variations in metallicity are 
less likely than variations in gravity for this sample of local field brown dwarfs.

$H-K$ appears to be an easy to obtain and straightforward indicator of gravity for
late T dwarfs. This is significant as, for a field sample with
a likely range in age of 1--5~Gyr (Dahn et al. 2002 estimate 2--4~Gyr
based on kinematic arguments), gravity corresponds directly to mass.  
This tight relationship is due to the small dependence of radius on age or
mass for brown dwarfs older than about 200~Myr (Burrows et al. 2001).
Figure 9 of Burrows et al. (1997) shows that log~$g=4.5$
implies a mass of 15~$\rm M_{Jupiter}$, log~$g=5.0$ a mass of
35~$\rm M_{Jupiter}$, and log~$g=5.5$ a mass of 75 $\rm
M_{Jupiter}$. We list the $H-K$ implied surface gravities for the later T
dwarfs in Table 11.

Recent investigations of spectroscopic gravity indicators for L and
T dwarfs include those by Lucas et al. (2001), Burgasser et al. (2003b),
Gorlova et al. (2003), Mart\'{\i}n \& Zapatero Osorio (2003) and
McGovern et al. (2004). Figure 2 of Mart\'{\i}n \& Zapatero Osorio shows 
synthetic spectra for $\rm T_{eff}=$1000~K from COND models by Allard et al. 
(2001).  The models imply that the lines of  K~I at 1.243 and 1.254 $\mu$m 
become weaker with increasing gravity for T dwarfs\footnote{Theoretically 
this is explained by the fact that the column
abundance of molecules above a given pressure level $P$ in an atmosphere
is proportional to $P/g$, where $g$ is the gravity.  In a higher gravity
atmosphere an outside observer must, all else being equal, look to higher
pressure to observe the same column of absorber as in a lower gravity
atmosphere.  The calculations of Lodders (1999) show that with rising
pressure at a fixed temperature chemical equilibrium increasingly favors
KCl over K.  Thus in a higher gravity atmosphere the total column of
potassium above the floor set by the continuum opacity is less than in a
lower gravity model.  The sensitivity to gravity should be greater in
later T (lower $\rm T_{eff}$) atmospheres since the line forming region
(roughly 1000 and 1500~K in the 1.15 and $1.25\,\rm \mu m$ regions
respectively)  falls closer to the chemical equilibrium boundary than for
earlier T dwarfs (see Figure 2 of Lodders 1999). In addition higher
pressures produce greater line broadening, thus decreasing the line depth.
}. Figure 7 shows $J$-band spectra of three T6($\pm0.5$) dwarfs
(SDSS J1110$+$0116, 2MASS J2339$+$13 and 2MASS J0937$+$2931) and two T8 dwarfs
(2MASS J0727$+$1710 and Gl 570D). These dwarfs span a range in $H-K$ color
and are identified in Figure 6. It can be seen that as $H-K$ increases
from top to bottom in Figure 7, the K~I lines strengthen, supporting the
interpretation of increasing $H-K$ as being due to decreasing gravity.  
The increase in K~I equivalent widths for these dwarfs is confirmed by the
higher resolution NIRSPEC data of McLean et al. (2003).  

Comparison of our Figure 7 with Figure 2 of Mart\'{\i}n \& Zapatero Osorio 
(2003) shows that the depths of the K~I lines seen in 2MASS J0937$+$2931 are
similar to the model predictions for log~$g \sim$5.5, and that the lines seen 
in SDSS J1110$+$0116 are almost as strong as the synthetic spectrum with
log~$g \sim$3.5.  Figure 6 and the models of Marley et al. (2002) imply that
SDSS J1110$+$0116 has log~$g$ between 4.0 and 4.5. There is as
yet no direct measurement of the effective temperature of this dwarf, but
the effective temperatures of other T6 dwarfs are in the range
900--1075~K (Golimowski et al. 2004).  If SDSS J1110$+$0116 has 
$\rm T_{eff}\approx 1000\,\rm K$ and $\log g \approx 4.2$, the evolutionary 
models of Burrows et al. (2003, their Figure 1) imply that it is a 10~$\rm
M_{Jupiter}$ brown dwarf with an age of about 1$\times$10$^8$ years, 
i.e. similar to that of the Pleiades cluster.  However we argue in 
\S  5.7.4 that one might shift the model contours on Figure 6 up and to the 
right to bring them into better agreement with observed $J-K$ and measured
effective temperatures.  In that case SDSS J1110$+$0116 would have a somewhat
larger gravity, mass, and age.  We thus adopt a more conservative estimated
mass of 10 -- 15$\,\rm M_{Jupiter}$ and an age of 1 -- 3 $\times10^8$ years.
We note also that the candidate young-cluster T dwarf S Ori 70 has an unusually
red $H-K_s$ color (Zapatero Osorio et al. 2002) apparently consistent with
$\log g < 4$.

%No evidence of recent star
%formation is seen in nearby areas of sky, which suggests that
%SDSSJ1110$+$0116 might be somewhat older and more massive.

\subsection{Absolute Magnitudes: The L/T Transition}

It has been apparent since the earliest discoveries of L and T dwarfs that
L-type objects evolve into T-type objects as they cool.  With decreasing 
effective temperatures the condensation level for the principle L dwarf 
condensates (iron and silicates) falls progressively deeper in the atmosphere. 
Unless upward mixing is very efficient, the clouds will eventually disappear 
beneath the photosphere, whose location is strongly wavelength-dependent for 
these objects. Both the general evolutionary cooling trend and the removal 
of condensates produce lower atmospheric temperatures; under these conditions 
the equilibrium chemistry rapidly begins to favor CH$_4$ over CO as the
dominant C-bearing species.  With less photospheric condensates to veil
the molecular bands and the growing importance of CH$_4$ opacity in the 
$K$-band, the objects turn to the blue in $J-K$.  Marley
(2000) employed a simple, one scale-height thick cloud layer to demonstrate 
that the sinking of a finite cloud deck explains the red to blue transition 
in $J-K$.  This was subsequently confirmed by models employing more elaborate
cloud models (Marley et al. 2002; Tsuji  2002; Allard et al. 2003).

The absolute magnitudes presented here can be used to better understand this
behavior.  Figures 8 and 9 show, respectively, absolute $J$ and $K$
magnitude against spectral type and $J-K$ color.  A fifth order polynomial
fit is shown to absolute magnitude against spectral type and the coefficients
of the fit are given in Table 12.  Known binaries were removed from the sample 
before fitting the data; the mean scatter around the fit is 0.4 magnitudes
for M$_J$ and 0.3 magnitudes for M$_K$. 

As noted by Burgasser et al. (2002b) these data suggest that the L to T 
dwarf transition may be more complex than implied by the picture of a
continuously sinking cloud.  The most notable discrepancy between the
simple picture and the data shown in these figures is the brightening seen
at $J$-band as the objects transition from L to T.  In this section we
summarize various suggested mechanisms to explain this behavior and
compare their predictions to the photometric data presented here.

\subsubsection{Thin Cloud Decks}

A finite-thickness cloud deck forming progressively 
lower in the atmosphere will eventually disappear from sight.  In the 
opposite extreme, dust that is well mixed through the entire observable 
atmosphere, as in the DUSTY models of the Lyon group (e.g. 
Allard et al. 2001), will by definition never disappear.  Models of such 
objects show that they simply become progressively redder as they cool 
and thus, due to veiling of the changing molecular bands, never exhibit an
L to T transition.

Although models with finite-thickness cloud decks do move from red,
L-like colors to blue, T-like colors, they tend to do so relatively
slowly.  This is because the cloud always has a finite thickness and thus
does not disappear from a given bandpass instantaneously.  During the time
the cloud is departing, say from $J$-band visibility, the overall
atmosphere is continuing to cool and become fainter.  Thus in a color
magnitude diagram (Figures 8 and 9) models with finite-thickness clouds
that are opaque enough to reach the colors of the latest L dwarfs
($J-K\sim2$) tend to leisurely turn to the blue as they cool and so
reach the colors of the early T dwarfs at too faint magnitudes.  The
$f_{\rm sed} = 3$ model in Figure 8 is an example.

Tsuji \& Nakajima (2003) proposed an interesting solution to this problem.  
They found a family of models with relatively thin cloud decks in which
the turnoff from red to blue in $J-K$ was a function of gravity.  In these
models low-gravity 10 $\rm M_{Jupiter}$ objects depart from what might be
called the ``L-type cooling sequence'' (the progressive reddening in $J-K$
with later spectral type) and turn from red to blue at a point almost 2
magnitudes brighter than  high-gravity 70 $\rm M_{Jupiter}$ objects.  A
similar, though less extreme, bright turn off is seen in the $f_{\rm
sed}=5$ family of models in our Figures 8 and 9. Tsuji \& Nakajima then suggest that
there is not a single evolutionary path in which objects first fade at $J$
band as they get redder and then brighten as they turn blue.  Rather they
propose that the brighter transition T dwarfs are low mass objects that
cooled to mid-L type and then turned from red to blue colors around $M_J
\sim 13.3$.  Dimmer transition objects would represent intermediate-mass
brown dwarfs that turned off the L cooling sequence at a later L type and
redder $J-K$, and and the latest Ls would represent the highest mass
objects.

This model makes a number of interesting predictions.  First, the $M_J$ vs.  
$J-K$ phase space between the L and T dwarfs should eventually be found to
be fairly evenly populated both at brighter and fainter magnitudes than
is shown by the transition objects detected to date.  Second, bright early T
dwarfs, like SDSS J1021$-$0304 (T3) or 2MASS J0559$-$1404 (T4.5) should be 
fairly low mass objects while the latest and reddest L dwarfs, like 2MASS 
J1632$+$1904 (L7.5), should be fairly high mass objects.  Emerging gravity 
indicators should be able to test this hypothesis.

\subsubsection{Patchy Clouds}

Plotted in the right panels of Figures 8 and 9 are model sequences from 
Marley et al. (2002).
Both cloud-free and cloudy models are shown, the latter with sedimentation
parameters $f_{\rm sed}=3$ and 5.  For the $f_{\rm sed}=5$ and the no-cloud
models, 3 gravities are shown ($\log g = 4.5$, 5, and 5.5). Only $\log g =
5$ is shown for $f_{\rm sed}=3$. Model effective temperatures are given on the
right axis of Figure 9.  The general agreement between the observed L
colors and the $f_{\rm sed}=3$ models seen in Figures 4, 8 and 9, 
suggests that the $JHK$ colors of the L dwarfs can be explained
by a uniform global cloud model.  The cloud's vertical extent and optical
depth are limited by sedimentation. Models with much less or much more
efficient sedimentation would be generally be too red or too blue,
respectively, than most of the L dwarf population.  However the existence
of a few dwarfs that are redder and a few that are bluer than
most (\S 5.5) implies that about 10\% of the L dwarf population would
require more extreme models.  These variations could arise from either
metallicity or sedimentation efficiency differences between objects.

However, the $f_{\rm sed} = 3$ models turn too slowly to
the blue and reach the colors of the bluest T dwarfs at $J$ band
magnitudes that are too faint. Faced with this difficulty of finite cloud
layers taking too long to disappear in models like those of Tsuji \&
Nakajima (2003), Burgasser et al. (2002b) propose a different mechanism for 
the L-to-T transition.  Drawing on a suggestion from Ackerman \& Marley 
(2001), Burgasser et al. (2002b)
propose that the L to T transition region is marked by the appearance of
holes in the global cloud deck, not unlike those seen in the ``5-$\rm \mu
m$ hot spots'' on Jupiter.  Deeply-seated flux, particularly in the clear
$Z$ and $J$ windows would then pour out of these holes, pushing the
disk-integrated color to the blue.  Indeed Burgasser et al. (2002b) found
that a relatively small fraction of holes would appreciably move a late 
L-type object towards the blue in $J-K$.  They argued that such a
mechanism would explain the brightening observed at $Z$ and $J$, and not
at other bands, across the transition, and also the observed resurgence 
in FeH absorption from the latest Ls to the early to mid Ts.

To illustrate the effect such holes might have,  we have joined with a dotted
line the magnitude:color values for the cloudy and cloud-free models at 
$\rm T_{eff}=1300$ K in both Figures 8 and 9.\footnote{Note that Burgasser
et al. presented a slightly more sophisticated interpolation,
computing colors using a weighted sum of flux produced by clear and cloudy
temperature profiles.  The straight lines in Figures 8 and 9 are
reasonable approximations to their ``clearing models''.}
The agreement between the datapoints for T1 to T3 dwarfs and the cloudy to 
cloud-free interpolations in Figures 8 and 9 is generally good.
The parameters  $\rm T_{eff}=1300$ K and $\log g =$ 4.5--5.5 appear
to bracket the known transition objects in these plots, in fair agreement with
Golimowski et al. (2004) who show that there is an apparent plateau at 
$\rm T_{eff}\approx$ 1450~K for types L7 to T4.
The cloud clearing model simply posits that at a given effective
temperature the global cloud deck begins to break up, perhaps because it
has settled sufficiently deeply into the convection zone that it becomes
subject to the global circulation pattern. The observed constancy of
$\rm T_{eff}$ across the transition is not required by this
hypothesis, although it does raise problems for the Tsuji \& Nakajima (2003)
suggestion of continuous cooling across the transition. 

If the clearing
does happen over a narrow temperature range, with a large spectroscopic
change occurring over a small change in temperature and luminosity as the
brown dwarf cools, we would not expect to discover many early T dwarfs.  
However about one-third of our T dwarf sample is made up of types T0--T3.5,
with a possible dearth of T3--4 types (Figure 3). A study of the
spectral type distribution in an SDSS magnitude-limited sample will
be the subject of a future paper (Collinge et al. 2002).

\subsubsection{Sudden Downpour}

Rather than relying on spatial inhomogeneities, Figures 8 and 9 suggests a 
third possibility for the L to T transition, which we term the ``sudden
downpour'' model.  The $f_{\rm sed} = 3$ models do a reasonably good job
of reproducing the colors of the latest L dwarfs.  Like the thin Tsuji \&
Nakajima (2003) cloud, models with more efficient sedimentation (larger 
$f_{\rm sed}$) turn off the L dwarf cooling track sooner (at brighter 
magnitudes) than seems to be consistent with the available data.  However,
one might argue that L dwarfs first cool at essentially constant $f_{\rm
sed}$, then at around $\rm T_{eff} = 1300\,\rm K$, $f_{\rm sed}$ begins to
gradually increase from $\sim 3$ to infinity at roughly fixed effective
temperature.  This rapid increase in the efficiency of sedimentation
would, in essence, produce a torrential rain of condensed iron and 
silicate grains.  Unlike the Tsuji \& Nakajima (2003) mechanism this would 
begin at the late L spectral type for all masses. T1 to T4 dwarfs would 
represent different stages of this cloud thinning process. Figures 8 and 9
shows where the $f_{\rm sed}=5$ models would lie, for example.  Once grains 
are essentially completely removed from the atmosphere the object would 
continue to evolve and cool.

A useful diagnostic for evaluating the various transition models may be
gravity. The sudden downpour mechanism, for example, would predict that
the T3.5 dwarf SDSS J1750$+$1759 (see Figure 8) would have $\log g \sim 
5.4$. Evolution tracks for the patchy cloud model curve downward at blue 
$J-K$ compared to the straight lines shown in the figure and thus this 
model would predict a smaller gravity, say $\log g = 5$.  On the other hand,
Tsuji \& Nakajima (2003) would predict that since this relative bright T 
dwarf has already made the transition to blue $J-K$, it must be relatively 
low in mass and have a substantially lower gravity, say $\log g = 4$.  There 
are also gravity tests among the late L dwarfs, although these are more 
subtle since all of the models would predict that the faintest and reddest L dwarfs 
will have progressively higher masses.  Tsuji \& Nakajima (2003) predict that 
the lowest mass dwarfs turn to the blue relatively early.  The downpour model 
also predicts that lower masses turn blueward earlier than higher masses as can
be seen by studying the tracks for the various gravities in the
$f_{\rm sed}=5$ case. However this turn off happens at later spectral types
than in the Tsuji \& Nakajima model and is accompanied by a subsequent
brightening in $M_J$.  For example the L7 dwarf labeled in Figure 8 could have
$\log g \ge 4.8$ under the cloud clearing and downpour models, but Tsuji
\& Nakajima (2003) would predict a higher minimum gravity, likely around 5.3 or
so. Certainly there are hints in Figure 8 that there is a width in $M_J$ to
the transition and this will facilitate such tests.  

There are also other diagnostics to consider.  The patchy-cloud model
straightforwardly accounts for the resurgence seen in FeH absorption
across the transition (Burgasser et al. 2002b), and it is not clear that
other models can account for this.  Clearly more modeling of all
mechanisms must be completed to better define gravity and other
diagnostics of the transition mechanism.  Unfortunately gravity indicators
among the late L and early T field dwarfs are elusive and it may be difficult 
to use them to definitively test the models. 
(One might expect that the Li test (Mart\'{\i}n, Rebolo \& Magazzu 1994) 
could be used to identify high gravity dwarfs since only 
brown dwarfs more massive than 0.06~M$_{\odot}$ will have
burned Li during their evolution.   As a practical matter, however, 
the Li line at 0.6708~$\mu$m in late L and early T dwarfs is detectable only 
with the largest telescopes due to the lack of continuum flux in this region.
Also, in T dwarf atmospheres LiCl and LiOH become the dominant Li bearing 
molecule (Lodders 1999) and these species are currently undetectable.)
Self consistent evolutionary models be developed as well as mechanisms to 
explain the onset of either patchiness or varying $f_{\rm sed}$
at a particular $\rm T_{eff}$.  We plan to more fully explore these issues
in a future paper.

\subsubsection{Additional Considerations}

Discrepant objects to note in Figures 8 and 9 are Kelu-1 (L3), SDSS J0423$-$0414
(T0) and 2MASS J0415$-$0935 (T9).  2MASS J0415$-$0935 is significantly redder in
$J-K$ than the models would predict, as discussed in \S 5.6.  Kelu-1 and SDSS 
J0423$-$0414 are both superluminous by about 0.75
magnitudes, suggesting that they may be pairs of identical dwarfs in
unresolved binary systems. Kelu-1 has been imaged by $HST$ and by
Keck, with no evidence of duplicity found (Mart\'{\i}n et al. 1999a;
Koerner et al. 1999), while SDSS J0423$-$0414 has not been imaged at high
resolution to our knowledge.  Cruz et al. (2003) classify this dwarf as
L7.5 using red spectra; however, CH$_4$ bands are clearly seen in the G02
spectrum (Figure 3 in G02), and the bolometric correction (i.e. ratio of
$K$-band flux to total luminosity) is more compatible with an infrared 
classification of T than L (Golimowski et al. 2004).  Although it has
been suggested that this object consists of a late L and early T close pair 
(Burgasser et al. 2003b) a discrepancy between the optical and infrared types 
is not unexpected.  As described in 
\S 5.2, different wavelength regions probe different levels of the photosphere.
It is likely that the optically derived spectral types are more representative 
of effective temperature, and as Golimowski et al. (2004) show that there is 
an apparent plateau at $\rm T_{eff}\approx$ 1450~K for types L7 to T4 we would 
expect optical types to be earlier than infrared types for the L/T transition
objects.  The earlier optical classification is in fact observed for six L8 
to T0 dwarfs in our sample, as indicated in Table 9.

Although the clear atmosphere models do a good job of reproducing the colors
of the later T dwarfs in the $J-H:H-K$ diagram (Figure 6), they tend to
predict bluer $J-K$ colors for these objects than is observed (Figures
8 and 9).  This seems to be a generic problem with clear models (see also Tsuji
\& Nakajima 2003).  The $J-K$ result implies that the model tracks shown in
Figure 6 should slide up and to the right, consistent with the suggestion in 
\S 5.6 that the temperature contours are too warm.  Because of the overall 
shape of the model contours the quality of the fit in $J-H:H-K$  would remain
about the same.  The discussion of the gravity signature seen in $H-K$ is still
qualitatively valid, although SDSS J1110$+$0116 would be expected to have a 
somewhat higher gravity.

Finally, we consider detectability limits for SDSS using
Figures 3, 8 and 9. Many SDSS T dwarfs have to be selected as $z$-only
objects and they are either not detected or only barely detected at $i$
(see Table~1).  The nominal (5$\sigma$, better than $1.5''$ seeing) $z$
detection limit is 20.8, and the faintest SDSS T dwarf discovered to date
is close to this limit, at $z$ = 20.4.  Since the colors of late (e.g. T8)
dwarfs are $z-J\approx$3.8, the corresponding $J$ limit is $\approx$16.6;
Table~9 therefore indicates that SDSS should be able to detect T9 dwarfs
to 10 pc.  The latest SDSS dwarf discovered to date is a T7, but we
anticipate that some later types will be found.

\section{Summary}

We have presented new near-infrared photometry and spectroscopy for cool
dwarfs from two sources: new very red objects from SDSS and known L and T
dwarfs from SDSS and 2MASS.  We have obtained new $JHK$ photometry for 71
L and T dwarfs (53 from SDSS and 18 from 2MASS), $Z$ photometry for 7 2MASS
objects, and spectroscopy of 56 L and T dwarfs (45 from SDSS and 11 from
2MASS).  The spectral types have been obtained using the classification
scheme of G02 which uses four molecular band indices at $J$, $H$ and
$K$.  The combined data from this and our previous papers are analyzed.  
Absolute magnitudes are available for 45 late-type dwarfs thanks to recent
parallax measurements.  The relationships among color, absolute magnitude
and spectral type are compared with model atmospheres with and without
clouds.  The major results and conclusions are as follows.

\begin{itemize}

\item

Of the 44 new SDSS targets for which infrared spectra were obtained,
one is an M dwarf, six are L dwarfs previously reported by Hawley et al. (2002) 
and Cruz et al. (2003) which are also classified as L in the G02 scheme, 
23 are new L dwarfs, and 14 are new T dwarfs (one of which was classified as 
late L from optical spectra by Hawley et al. 2002).  The new T dwarf sample
consists of seven T0--T2 and seven T4.5--7 types, and we also identify nine L
dwarfs with type L8 and later.  These observations add significantly to the
sample of L/T transition objects, and bring to 58 the total number of published 
T dwarf systems.

\item

We provide provisional indices on the G02 scheme for the end of the T
spectral sequence. The spectral type of 2MASS J0415$-$0935 is T9;  it is
currently the coolest known dwarf, with an effective temperature of $\sim$
700~K (Golimowski et al. 2004, Vrba et al. 2004).

\item

As recognized previously, the relatively muted colors of the L dwarfs
(compared to models in which there is no sedimentation) imply that
silicate and iron cloud optical depths are limited by condensate
sedimentation.

\item 

As noted in previous work, the $JHK$ colors of mid to late L dwarfs show 
a large scatter within a given spectral type.  
The $JHK$ colors are reasonably reproduced by models which incorporate cloud
formation with a modest range of condensate sedimentation efficiencies (or
equivalently cloud optical depth).  About 10\% of the dwarfs in our sample
seem to either be substantially bluer or redder than a modest range in
$f_{\rm sed}$, of about 3 to 4, would predict.  This suggests that cloud
properties are generally similar, but can differ, among L dwarfs.

\item 

The reddest and bluest L dwarfs show a scatter in the infrared spectral 
indices, and there can also be significant differences between the spectral 
types determined using optical and near-infrared spectra.  The differences 
between indices can be understood in terms of the depths probed by the 
different wavelength regions. 

\item

The near-infrared colors of T dwarfs become rapidly bluer towards later
types. However, beyond about T5, these colors, especially $H-K$, show
large scatter.  This is correlated with the equivalent width of the
$J$-band K~I doublet absorption, strongly suggesting that the $H-K$ color
is gravity-dependent.  Model atmospheres show that the $K$-band flux is
depressed by $\rm H_2$ absorption and that at a given effective
temperature $H-K$ becomes bluer with increasing gravity.  As intermediate
age T dwarfs all have essentially the same radius, $H-K$ is a good
indicator of mass for field T dwarfs of a given spectral type.

\item

The implied masses of almost all of the T dwarfs in the sample are 15 -- 75
$\rm M_{Jupiter}$. The T5.5 dwarf SDSS J1110$+$0116 appears to have a particularly
low mass (10 -- 15 $\rm M_{Jupiter}$) with an inferred age of about
1 -- 3$\times10^8$ years.

\item

The absolute magnitude--spectral type relationship for L and T dwarfs
shows a mostly steady decline towards later spectral type, from $M_J$
$\sim$ 11.5, $M_K$ $\sim$ 10.5 at L0 to $M_J$ $\sim$ 16.5, $M_K$ $\sim$ 17
at T9.  There is a peak or plateau, depending on wavelength, in absolute
magnitude between types L7 and T4 (where $\rm T_{eff}\approx$1450~K,
Golimowski et al. 2004).  
Models (e.g. Burgasser et al. 2002b) which invoke an onset of some type of
modification to the vertical or horizontal extent of the cloud at a fixed
$\rm T_{eff}$ seem to better explain this observation than models which
assume continuous sinking of a cloud that is spatially and vertically
uniform over time.  Additional observational tests of the various cloud
disruption possibilities are needed.

\end{itemize}

\acknowledgments 

We are most grateful to the staff at UKIRT for their assistance in
obtaining the data presented in this paper.  Some data were obtained
through the UKIRT Service Programme.  UKIRT is operated by the Joint
Astronomy Centre on behalf of the U.~K.\ Particle Physics and Astronomy
Research Council.   DAG thanks the Center for Astrophysical Sciences at 
Johns Hopkins University for its moral and financial support of this work.
GRK is grateful for support to Princeton University and
to NASA via grants NAG5-8083 and NAG5-11094.  MSM acknowledges support
from NASA grants NAG2-6007 and NAG5-8919 and NSF grant AST 00-86288.
TRG's research is supported by the Gemini Observatory,
which is operated by the Association of Universities for Research in
Astronomy on behalf of the international Gemini partnership of Argentina,
Australia, Brazil, Canada, Chile, the United Kingdom, and the United
States of America.
Funding for the creation and distribution of the SDSS Archive has been
provided by the Alfred P. Sloan Foundation, the Participating
Institutions, the National Aeronautics and Space Administration, the
National Science Foundation, the U.A. Department of Energy, the Japanese
Monbukagakusho, and the Max Planck Society.  The SDSS web site is {\bf
http://www.sdss.org}. The SDSS is managed by the Astrophysical Research
Consortium (ARC) for the Participating Institutions.  The Participating
Institutions are the University of Chicago, Fermilab, the Institute for
Advanced Study, the Japan Participation Group, The Johns Hopkins
University, Los Alamos National Laboratory, the Max-Planck-Institute for
Astronomy (MPIA), the Max-Planck-Institute for Astrophysics (MPA), New
Mexico State University, University of Pittsburgh, Princeton University,
the United States Naval Observatory, and the University of Washington.

\newpage 
\begin{figure}
\epsscale{0.3}
%\plotone{/home/skl/data/skl/sdss/2002/jhkpaper/charts1.ps}
\plotone{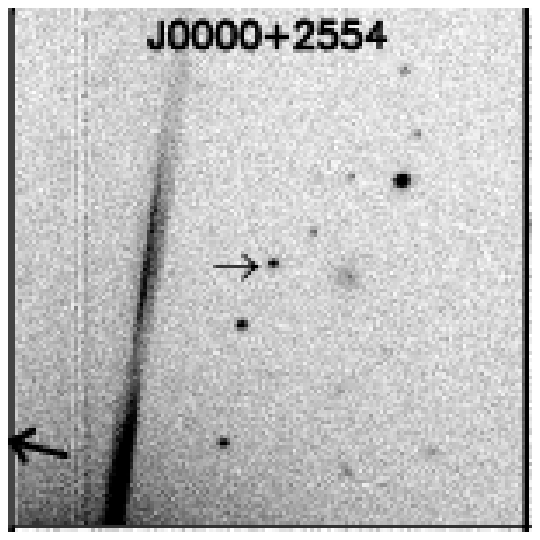}
\end{figure}
\newpage
\begin{figure}

%\plotone{/home/skl/data/skl/sdss/2002/jhkpaper/charts2.ps}
%\plotone{knapp.fig1b.ps}
\caption{The SDSS $z$ finding charts of the new dwarfs listed in Table 1.
The side of each finding chart is 120$''$. The large direction arrow points north, with east $90^{\circ}$ 
counterclockwise from north. The object is indicated by the small arrow.
[THUMBNAIL ONLY FOR COMPLETE CHART SEE JPEGS UNDER POSTSCRIPT OPTION.]
\label{fig1}}
\end{figure}

\newpage
\begin{figure}
\plotfiddle{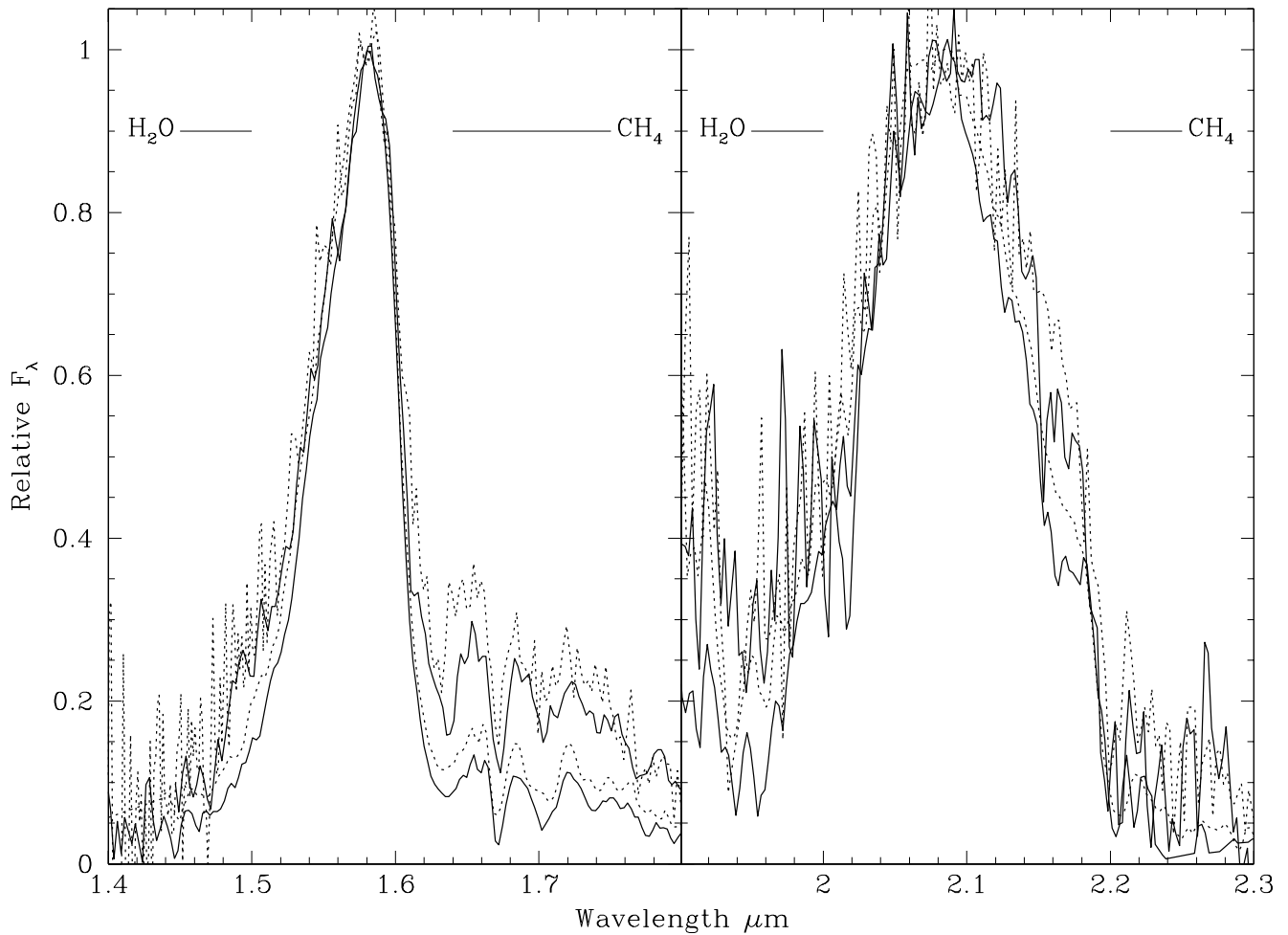}{14truecm}{-90}{70}{70}{-270}{440}
%\plotfiddle{/home/skl/data/skl/sdss/2002/jhkpaper/HK_sp_T.ps}{14truecm}{-90}{70}{70}{-270}{440}
%%\plotfiddle{knappHKspT.ps}{14truecm}{-90}{70}{70}{-270}{440}
\caption{Normalized $H$ and $K$ spectra (R$\approx$400) 
for, from top to bottom, SDSS J1624$+$0029 (T6), 
SDSS J1758$+$4633 (T7), Gl 570D (T8) and 2MASS J0415$-$0935 (T9).  
The spectra for SDSS J1624+0029 (Strauss et al. 1999)
and Gl 570D (Geballe et al. 2001) are shown as dotted lines.
\label{fig2}}
\end{figure}
\newpage

\clearpage
\begin{figure}
\plotfiddle{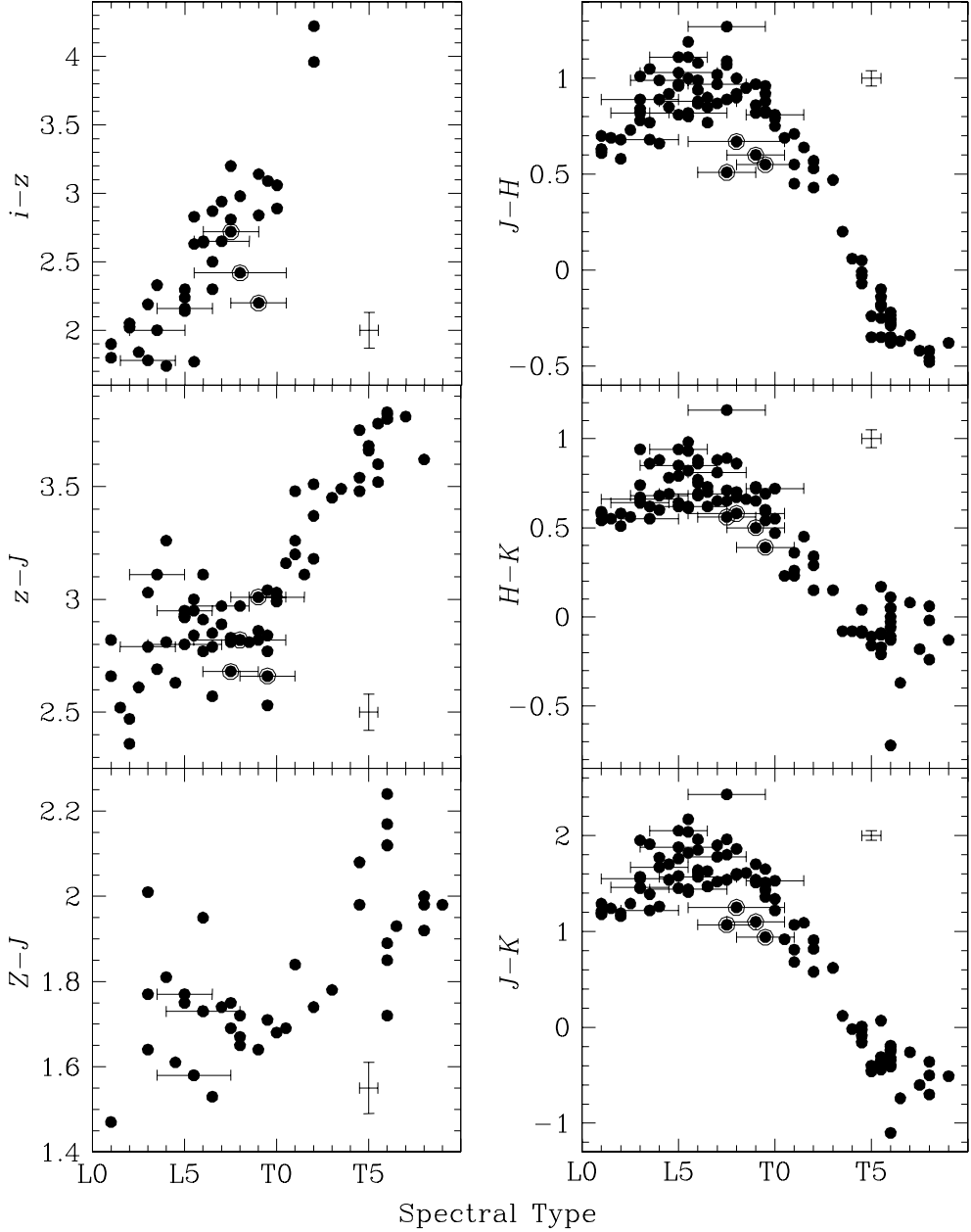}{17truecm}{0}{70}{70}{-240}{-50}
%\plotfiddle{/home/skl/data/skl/sdss/2002/jhkpaper/izZJHK_type.ps}{17truecm}{0}{70}{70}{-240}{-50}
%%\plotfiddle{knappcol.ps}{17truecm}{0}{70}{70}{-240}{-50}
\caption{Colors as a function of spectral type.  Note that SDSS $i$ and $z$ 
measurements are on the AB system, while the other magnitudes are on the 
Vega $=$ 0 system.  $Z$ is on the UKIRT (UFTI) photometric system while $JHK$ 
are on the MKO-NIR system. Points of low signal-to-noise ratio are not plotted.  
Typical error bars are shown as well as those for dwarfs with spectral type 
uncertainty larger than one subclass.  Data points for the four unusually blue 
late L dwarfs are ringed (\S 5.5).
\label{fig3}}
\end{figure}
\newpage
\clearpage

\begin{figure}
\plotfiddle{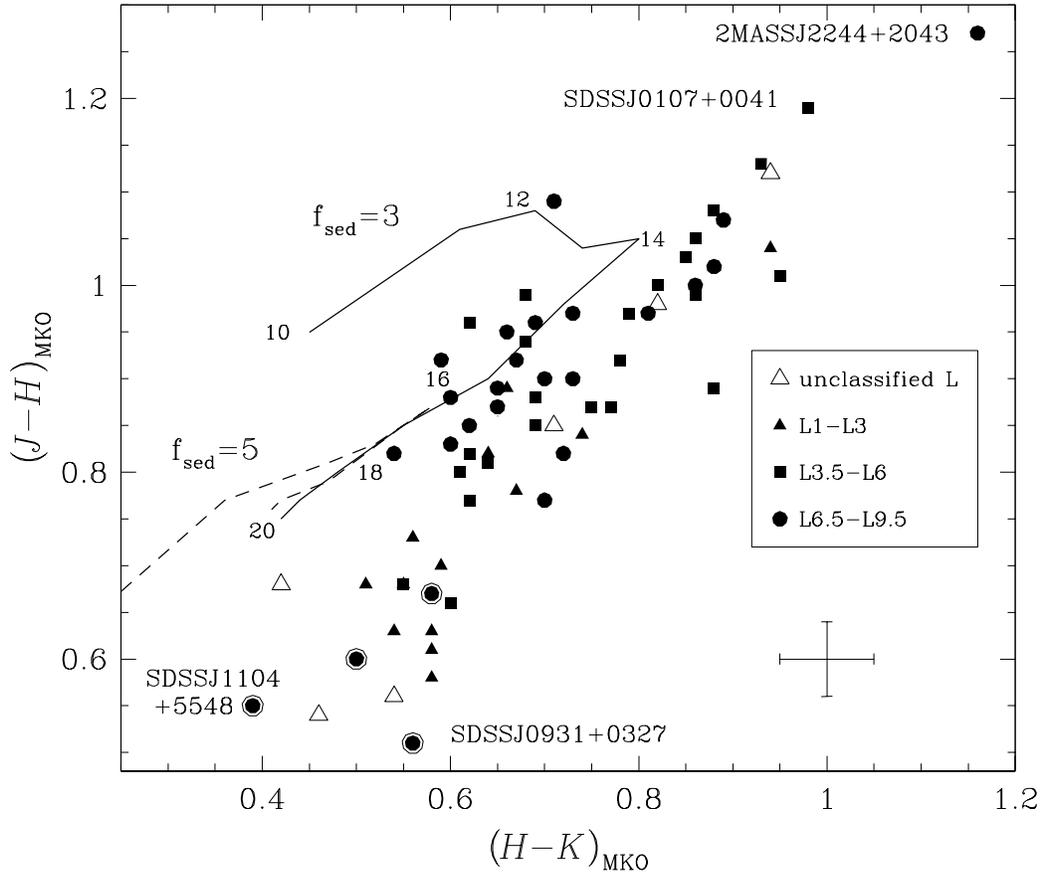}{14truecm}{0}{70}{70}{-250}{0}
%\plotfiddle{/home/skl/data/skl/sdss/2002/jhkpaper/jhk_L.ps}{14truecm}{0}{70}{70}{-250}{0}
%%\plotfiddle{knappjhkL.ps}{14truecm}{0}{70}{70}{-250}{0}
\caption{$J-H$ vs. $H-K$ for the L dwarfs in Table 9.  Model sequences
(Marley et al. 2002) are shown for $f_{\rm sed}=3$ (solid line) and 
$f_{\rm sed}=5$ (dashed line),
log~$g=$~5.0, cloudy atmospheres.  $\rm T_{eff}$ is indicated in units of
100 K along the $f_{\rm sed}=3$ sequence. Dwarfs with extreme colors are
identified, data points for four unusually blue late L dwarfs are ringed. 
A typical error bar is shown.
\label{fig4}}
\end{figure}
\newpage

\begin{figure}
\plotfiddle{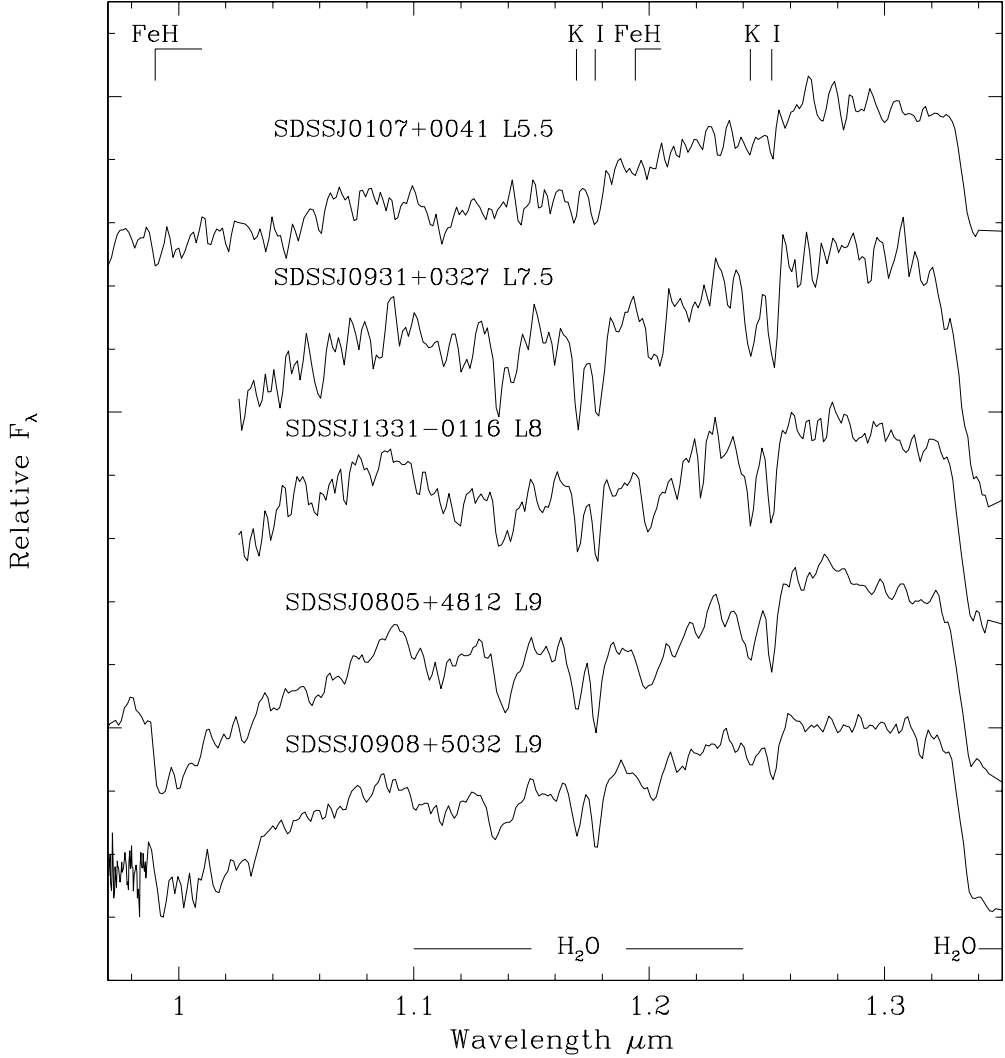}{14truecm}{0}{70}{70}{-250}{0}
%\plotfiddle{/home/skl/data/skl/sdss/2002/jhkpaper/J_blueL.ps}{14truecm}{0}{70}{70}{-250}{0}
%\plotfiddle{knappJblueL.ps}{14truecm}{0}{70}{70}{-250}{0}
\caption{$J$-band spectra (R$\approx$600) for three of the four unusually blue late L dwarfs, bracketed by more typical L5.5 and L9 dwarfs.  We identify the strong FeH 
features at 0.99 and 1.19~$\mu$m, K~I lines at 1.169, 1.177, 1.243 and 1.252~$\mu$m, and $\rm H_2O$ bands around 1.15 and 1.33~$\mu$m.
See Cushing et al. (2003) for a more complete identification of FeH features.
\label{fig5}}
\end{figure}
\newpage

\begin{figure}
\plotfiddle{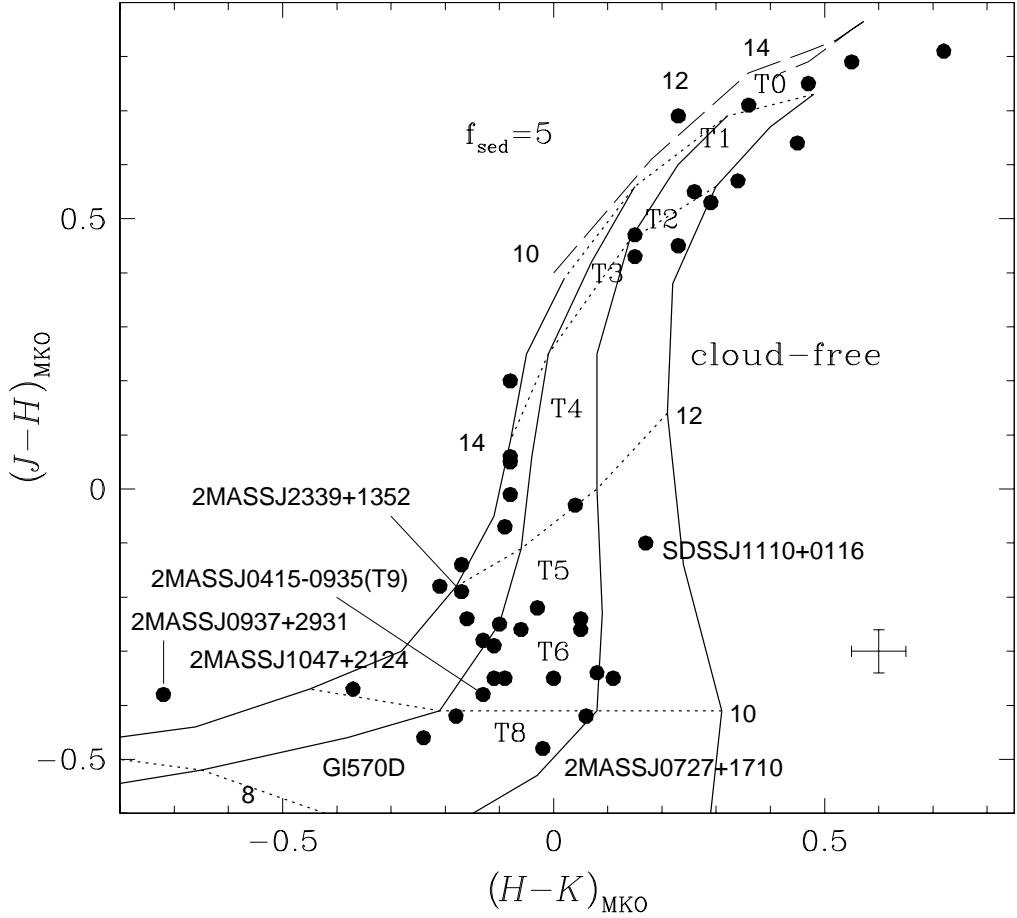}{14truecm}{0}{70}{70}{-250}{0}
%\plotfiddle{/home/skl/data/skl/sdss/2002/jhkpaper/jhk_T.ps}{14truecm}{0}{70}{70}{-250}{0}
%\plotfiddle{knappjhkT.ps}{14truecm}{0}{70}{70}{-250}{0}
\caption{$J-H$ vs. $H-K$ for the T dwarfs in Table 9.  Model sequences
from Marley et al. (2002) are shown for $f_{\rm sed}=5$ and log~$g=$~5.0
cloudy models (dashed line) and cloud-free models (solid line)  where
from left to right log~$g=$~5.5, 5.0, 4.5 and 4.0.  $\rm T_{eff}$ is
indicated in units of 100 K and for the cloud-free models dotted lines
indicate constant $\rm T_{eff}$.  The typical error is shown at the bottom
right, and the sequence of spectral type is shown. Dwarfs of particular
interest are labelled (see \S 5.6).  For field T dwarfs, gravity correlates
with mass, and the range in log~$g$ shown here translates into masses 75,
35, 15 and 5 $\rm M_{Jupiter}$ from left to right.
\label{fig6}}
\end{figure}
\newpage

\begin{figure}
\epsscale{0.9}
\plotone{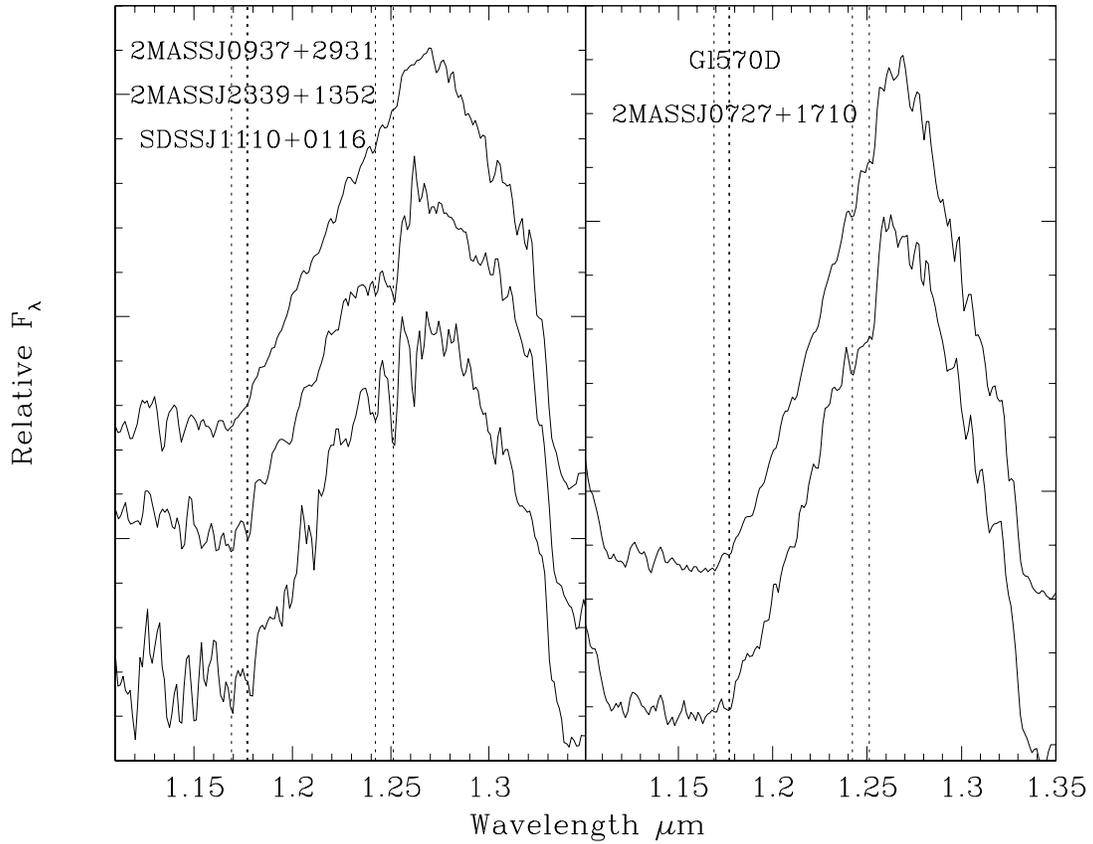}
%\plotone{/home/skl/data/skl/sdss/2002/jhkpaper/J_sp.ps}
%\plotone{knappJsp.ps}
\caption{ Normalized $J$-band spectra (R$\approx$600) for a sample of 
T6($\pm0.5$) dwarfs (left) and T8 dwarfs (right).  The location of K~I 
doublets are shown by the dotted lines. The dwarfs have $H-K$ increasing 
from top to bottom, and the K~I features strengthen from top to bottom,
supporting the conclusion that log~$g$ decreases from top to bottom.
\label{fig7}}
\end{figure}
\newpage

\begin{figure}
\plotfiddle{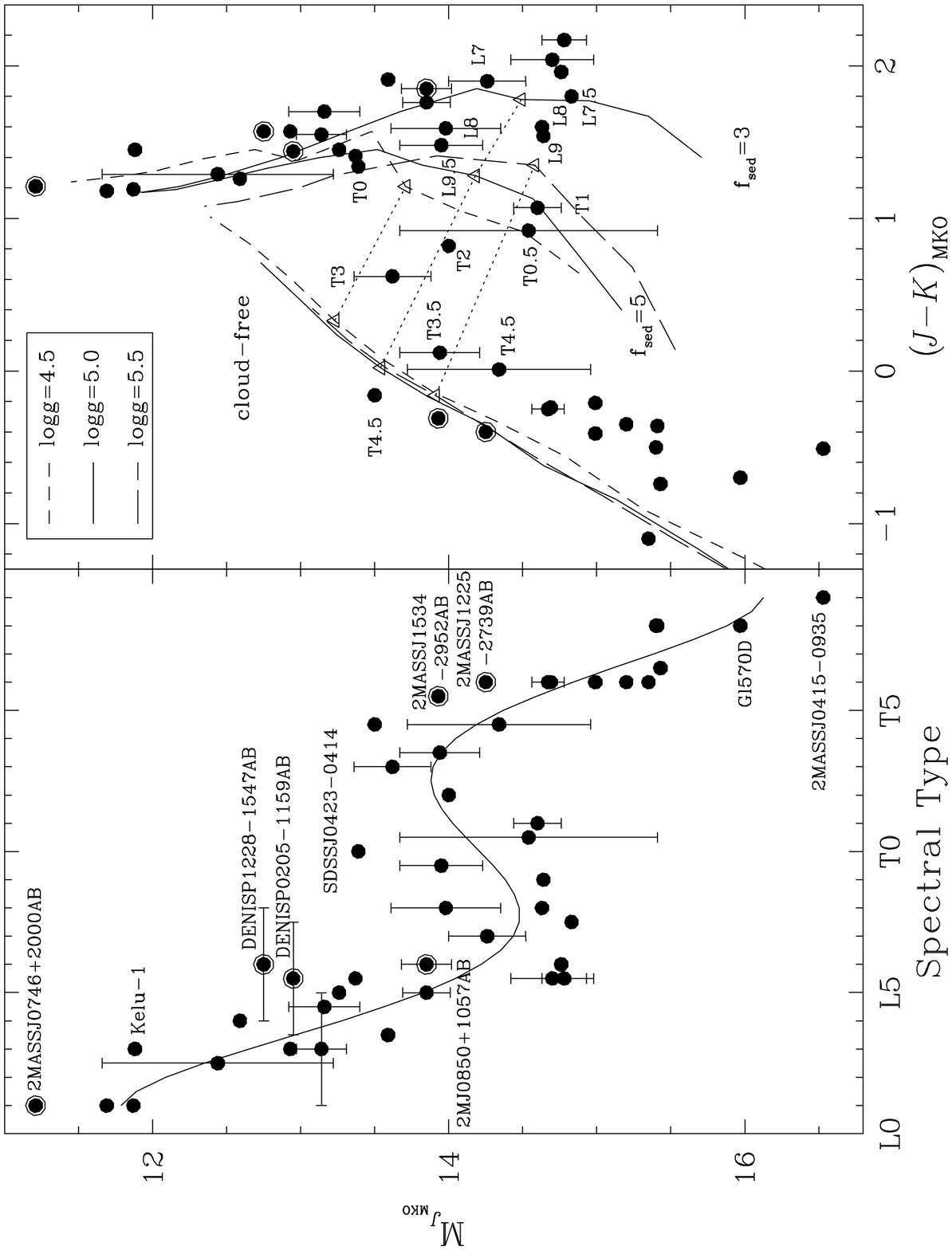}{14truecm}{-90}{70}{70}{-300}{450}
%\plotfiddle{/home/skl/data/skl/sdss/2002/jhkpaper/mj_logg.ps}{14truecm}{-90}{70}{70}{-300}{450}
%\plotfiddle{knappmjlogg.ps}{14truecm}{-90}{70}{70}{-300}{450}
\caption{M$_J$ as a function of spectral type and $J-K$ color. Error bars 
are shown where the distance modulus is uncertain by $\geq$~0.1~mag and 
where type is uncertain by $>$1 subclass. Known binaries are indicated by 
ringed symbols.  In the left panel the known binaries and other apparently 
superluminous dwarfs are labelled, as well as the two faintest dwarfs.  A 
5th order polynomial fit to $M_J$:type is shown (coefficients are given in 
Table 12).  In the right panel all objects with types between L7 and T4.5 
are identified. Model $M_J$ against $J-K$ sequences are shown for 
$f_{\rm sed}=3$, $f_{\rm sed}=5$ and cloud-free models from Marley et al. 
(2002). The dotted lines between the triangles connect the $\rm T_{eff}=1300$~K 
points on each model sequence (see discussion in \S 5.7).
\label{fig8}}
\end{figure}
\newpage

\begin{figure}
\plotfiddle{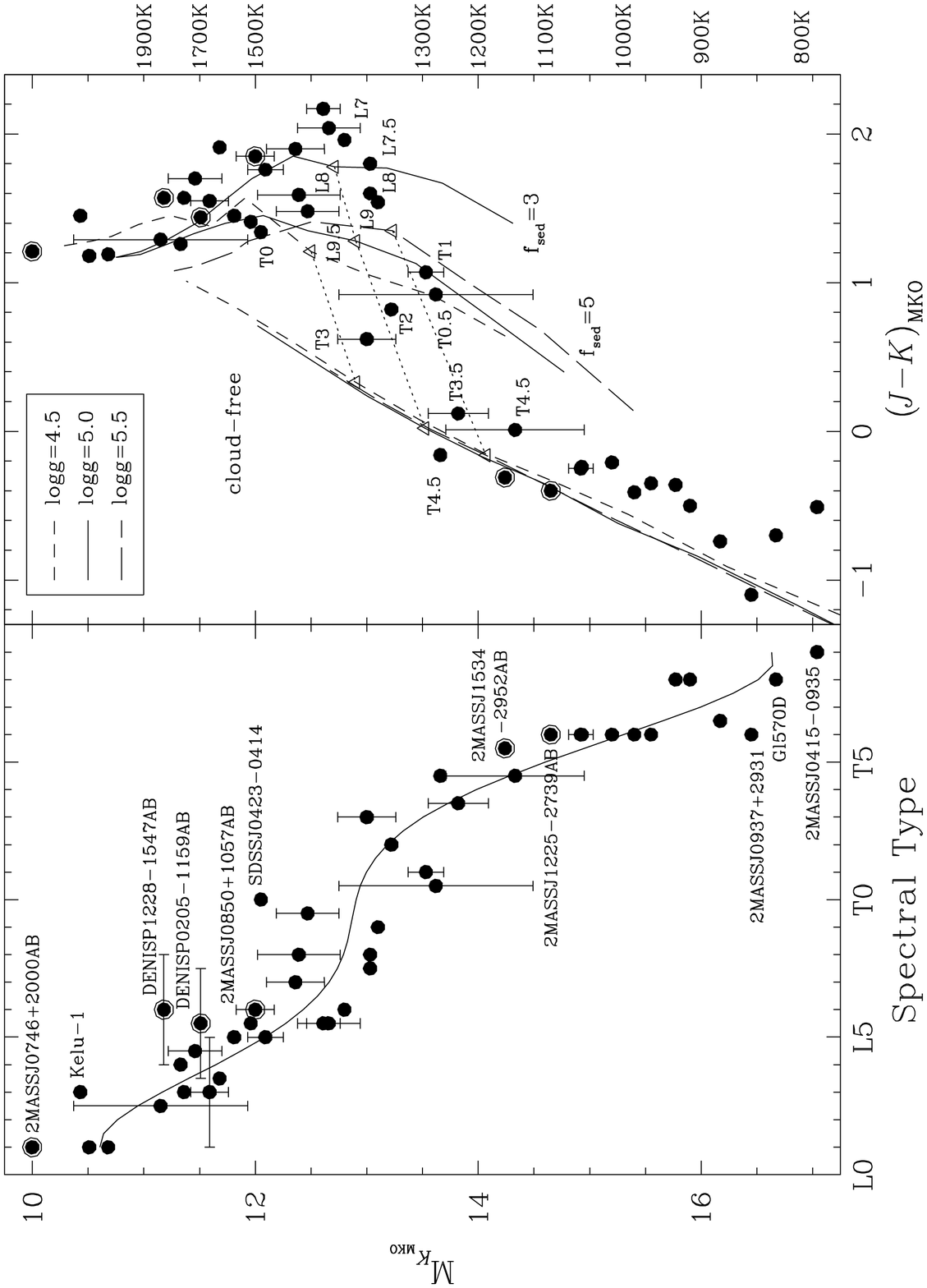}{14truecm}{-90}{70}{70}{-300}{450}
%\plotfiddle{/home/skl/data/skl/sdss/2002/jhkpaper/mk_logg.ps}{14truecm}{-90}{70}{70}{-300}{450}
%\plotfiddle{knappmklogg.ps}{14truecm}{-90}{70}{70}{-300}{450}
\caption{M$_K$ as a function of spectral type and $J-K$ color. Error bars are 
shown where the distance modulus is uncertain by $\geq$~0.1~mag and where type 
is uncertain by $>$1 subclass. Known binaries are indicated by ringed symbols.
In the left panel the known binaries and other apparently superluminous dwarfs 
are labelled as well as the three faintest dwarfs.  A 5th order polynomial fit to 
$M_K$:type is shown (coefficients are given in Table 12). In the right panel all 
objects with types between L7 and T4.5 are identified. Model $M_K$
against $J-K$ sequences are shown for $f_{\rm sed}=3$, $f_{\rm sed}=5$ and
cloud--free models from Marley et al. (2002).  The dotted lines between
the triangles connect the $\rm T_{eff}=1300$~K points on each model
sequence. Effective temperatures from the log~$g=$~5.0 models are shown on
the right axis, where $\rm T_{eff}~=~$1500--1900~K correspond to the $f_{\rm
sed}=3$ models and $\rm T_{eff}~=~$800--1300~K correspond to the cloud-free
models.
\label{fig9}}
\end{figure}

\clearpage
\begin{deluxetable}{lrrrr}
\small
\tablenum{1}
\tablewidth{350pt}
\tablecaption{New Confirmed Late-M, L, and T Dwarfs from SDSS}
\tablehead{
\colhead{Object} &  \colhead{$i$}&
\colhead{$\sigma_i$}& \colhead{$z$}& 
\colhead{$\sigma_z$} \nl
}
\tablecolumns{5}
\startdata 

SDSS J000013.54$+$255418.6 & 25.59 & 0.56 & 18.48 & 0.04 \nl
SDSS J000112.18$+$153535.5 & 20.29 & 0.04 & 18.55 & 0.03  \nl
SDSS J001608.44$-$004302.3 & 21.11 & 0.11   & 19.34 & 0.07  \nl 
SDSS J012743.50$+$135420.9\tablenotemark{a}&  22.27  & 0.17 &  19.62 &  0.07 \nl 
SDSS J020333.26$-$010812.5& 22.76 &  0.36  & 20.36  & 0.15  \nl
\nl
SDSS J020735.60$+$135556.3\tablenotemark{a} &19.84  & 0.03 & 18.06  & 0.02  \nl
SDSS J035448.73$-$002742.1 & 21.63 &  0.10 &  19.61 &  0.07 \nl
SDSS J040100.96$-$060933.0 & 22.93  & 0.49 &  20.19  & 0.20 \nl
SDSS J074007.30$+$200921.9&  22.91  & 0.34  & 19.78  & 0.08 \nl 
SDSS J074149.15$+$235127.5&  24.88  & 0.39  & 19.65  & 0.05 \nl
\nl
SDSS J074201.41$+$205520.5& 23.85 &  0.75  & 19.28 &  0.05 \nl
SDSS J074719.71$+$293748.6&  21.64  & 0.10 &  19.91  & 0.08 \nl
SDSS J075515.26$+$293445.4 & 21.73 &  0.18  & 19.40 &  0.07 \nl
SDSS J075656.54$+$231458.5 &  22.89  & 0.31  & 19.82  & 0.07 \nl
SDSS J075840.33$+$324723.4 &  21.92  & 0.13  & 17.96  & 0.03 \nl
\nl
SDSS J080531.80$+$481233.0\tablenotemark{a} & 19.82  & 0.05  & 17.62 &  0.03 \nl
SDSS J080959.01$+$443422.2 & 21.82 &  0.16  & 19.28  & 0.06 \nl 
SDSS J083048.80$+$012831.1 &  24.91 &  0.52  & 19.59  & 0.08 \nl
SDSS J083120.81$+$304417.1 & 21.45 &  0.11  & 19.68  & 0.10  \nl
SDSS J085234.90$+$472035.0 & 21.99 &  0.15  & 18.90  & 0.05  \nl
\nl
2MASS J090837.97$+$503208.0\tablenotemark{b} &  20.06  & 0.03 &  17.22 &  0.02 \nl
SDSS J093109.56$+$032732.5&  22.00 &  0.15 &  19.28  & 0.05  \nl
SDSS J100401.41$+$005354.9 &  21.81 &  0.18 &  19.76  & 0.11 \nl
SDSS J103026.78$+$021306.4 &  23.43  & 0.61 &  19.94  & 0.11 \nl
SDSS J104409.43$+$042937.6 & 21.67  & 0.08  & 18.73  & 0.03  \nl
\nl
SDSS J104625.76$+$042441.0 &  22.38  & 0.16 &  19.74 &  0.07 \nl
SDSS J110454.25$+$554841.4 & 22.95 &  0.34 &  19.94 &  0.09 \nl
SDSS J112615.25$+$012048.2  &  22.28  & 0.25  & 19.79 &  0.12\nl
SDSS J115553.86$+$055957.5&  21.26 &  0.12 &  18.45 &  0.04 \nl
SDSS J115700.50$+$061105.2 & 24.12 &  0.67 &  20.20  & 0.11  \nl
\nl
SDSS J120747.17$+$024424.8\tablenotemark{a} &  21.47 &  0.11  & 18.41  & 0.04 \nl
SDSS J121001.96$+$030739.2&  22.25 &  0.26  & 19.79 &  0.10 \nl
SDSS J123147.39$+$084730.7 &  22.79 &  0.30  & 18.94 &  0.04 \nl
SDSS J133148.90$-$011651.4\tablenotemark{a}  &20.56 &  0.08  & 18.14 &  0.04 \nl
SDSS J135923.99$+$472843.2 &23.15  & 0.37  & 19.76 &  0.09  \nl
\nl
SDSS J140814.74$+$053952.9\tablenotemark{a} &  20.48 &  0.05 &  18.71 &  0.05 \nl
SDSS J143211.74$-$005900.8 &  22.00  & 0.23  & 19.62 &  0.10 \nl
SDSS J143535.70$-$004347.0\tablenotemark{a} &  20.86&  0.06& 19.02 &  0.04 \nl
SDSS J144016.20$+$002638.9 & 20.65  & 0.05  & 18.75  & 0.03  \nl
SDSS J151603.03$+$025928.9 &  22.25  & 0.26  & 19.89 &  0.12 \nl
\nl
SDSS J152103.24$+$013142.7 &  24.46  & 0.57  & 19.57 &  0.06 \nl
SDSS J152531.32$+$581053.1 & 22.05 &  0.16  & 19.75 &  0.09  \nl
SDSS J161626.46$+$221859.2 & 23.07  & 0.26 &  20.33 &  0.10  \nl
SDSS J162441.00$+$444145.8 & 23.17 &  0.40  & 20.07 &  0.11 \nl
SDSS J163030.35$+$434404.0 &  22.10  & 0.14 &  19.45 &  0.05 \nl
\nl
SDSS J163239.34$+$415004.3 &  23.29 &  0.37 &  20.35 &  0.11 \nl
SDSS J164939.34$+$384248.7 &  22.36 &  0.31 &  20.05 &  0.10 \nl
SDSS J175024.01$+$422237.8 &  24.07  & 0.96 &  19.38 &  0.09 \nl
SDSS J175805.46$+$463311.9 &  24.18  & 0.57  & 19.67 &  0.07 \nl
SDSS J204749.61$-$071818.3 & 23.87 &  0.78  & 19.74 &  0.10  \nl
\nl
SDSS J212413.89$+$010000.3 & 23.77 & 0.54 & 19.71 & 0.12 \nl

\tablenotetext{a}{Optical spectroscopy given by Hawley et al. (2002).}
\tablenotetext{b}{Discovered in the 2MASS database by Cruz et al. (2003). }
\tablenotetext{}{The SDSS asinh magnitudes (Lupton et al. 1999) corresponding 
to zero flux are $i=$24.4 and $z=$22.8.}

\enddata

\end{deluxetable}

%\newpage

\begin{deluxetable}{lrc}
\small
\tablenum{2}
\tablewidth{250pt}
\tablecaption{New UFTI $Z$ Photometry}
\tablehead{
\colhead{Name} &  \colhead{$Z$} &  \colhead{Date}  \nl 
\colhead{} &  \colhead{$\pm 0.05$} &  \colhead{(YYYYMMDD)} \nl 
\colhead{} &  \colhead{(mag)} &  \colhead{} \nl
}
\tablecolumns{3}
\startdata 
2MASS J0030$-$1450 &  18.13  & 20021106 \nl 
2MASS J0243$-$2453 &  16.98  & 20021106 \nl 
2MASS J0415$-$0935 &  17.30  & 20021106 \nl 
2MASS J0727$+$1710 &  17.17  & 20021106 \nl 
2MASS J0825$+$2115 &  16.62  & 20021106 \nl 
2MASS J0937$+$2931 &  16.01  & 20021106 \nl 
2MASS J2356$-$1553 &  17.60  & 20021022 \nl 
\enddata

\end{deluxetable}

\newpage

\begin{deluxetable}{lrrrrrrcc}
\small
\tablenum{3}
%\tablewidth{400pt}
\tablecaption{New  MKO--NIR $JHK$ Photometry}
\tablehead{
\colhead{Name}& \colhead{$J$}& \colhead{$\sigma_J$}& \colhead{$H$}& 
\colhead{$\sigma_H$}
& \colhead{$K$} & \colhead{$\sigma_K$} & \colhead{Date} & \colhead{Camera} \nl
\colhead{} & \colhead{} & \colhead{} & \colhead{} & \colhead{}
& \colhead{} & \colhead{} & \colhead{YYYYMMDD}& \colhead{}  \nl
}
\tablecolumns{9}
\startdata 
2MASS J0028$+$1501& 16.45& 0.03& 15.42& 0.03& 14.51& 0.03& 20011124& IRCAM \nl
2MASS J0028$+$1501& 16.44& 0.03& 15.44& 0.03& 14.53& 0.03& 20020108& UFTI\nl
2MASS J0030$-$1450& 16.39& 0.03& 15.37& 0.03& 14.49& 0.03& 20011124& IRCAM\nl
2MASS J0036$+$1821& 12.29& 0.03& 11.65& 0.03& 11.05& 0.03& 20021207& UIST\nl
2MASS J0243$-$2453& 15.13& 0.03& 15.39& 0.03& 15.34& 0.03& 20011124& IRCAM\nl
\nl
2MASS J0415$-$0935& 15.32& 0.03& 15.70& 0.03& 15.83& 0.03& 20010829& UFTI\nl
2MASS J0727$+$1710& 15.19& 0.03& 15.67& 0.03& 15.69& 0.03& 20011125& IRCAM\nl
2MASS J0755$+$2212& 15.46& 0.03& 15.70& 0.03& 15.86& 0.03& 20020109& UFTI\nl
2MASS J0801$+$4628&  16.21& 0.03& 15.31& 0.03& 14.58& 0.03& 20020109& UFTI\nl
2MASS J0908$+$5032&  14.40& 0.03& 13.54& 0.03& 12.89& 0.03& 20020108& UFTI\nl
\nl
2MASS J0937$+$2931& 14.29& 0.03& 14.67& 0.03& 15.39& 0.06& 20011124& IRCAM\nl
2MASS J1503$+$2525& 13.55& 0.03& 13.90& 0.03& 13.99& 0.03& 20030104& UIST \nl
2MASS J1534$-$2952AB& 14.60& 0.03& 14.74& 0.03& 14.91& 0.03& 20020715& IRCAM\nl
2MASS J1553$+$1532AB& 15.34& 0.03& 15.76& 0.03& 15.95& 0.03& 20020109& UFTI\nl
2MASS J2224$-$0158& 13.89& 0.03& 12.84& 0.03& 11.98& 0.03& 20020620& UFTI\nl
\nl
2MASS J2244$+$2043& 16.33& 0.03& 15.06& 0.03& 13.90& 0.03& 20020620& UFTI\nl
2MASS J2254$+$3123& 15.01& 0.03& 14.95& 0.03& 15.03& 0.03& 20010829& UFTI\nl
2MASS J2339$+$1352& 15.81& 0.03& 16.00& 0.03& 16.17& 0.03& 20010829& UFTI\nl
2MASS J2356$-$1553& 15.48& 0.03& 15.70& 0.03& 15.73& 0.03& 20011124& IRCAM\nl
SDSS J0000$+$2554 & 14.73 & 0.05 & 14.74 & 0.03 & 14.82 & 0.03 & 20031207 & UFTI \nl
\nl
SDSS J0001$+$1535 & 15.29 & 0.03 & 14.40 & 0.03 & 13.52 & 0.05 & 20031207 & UFTI \nl
SDSS J0016$-$0043 & 16.34 & 0.05 & 15.34 & 0.05 & 14.52 & 0.03 & 20031207 & UFTI \nl 
SDSS J0127$+$1354&  16.71& 0.03& 15.84& 0.03& 15.09& 0.03& 20020108& UFTI\nl
SDSS J0203$-$0108&  17.83& 0.05& 16.87& 0.03& 16.18& 0.03& 20030104& UIST\nl
SDSS J0207$+$1355&  15.27& 0.03& 14.45& 0.03& 13.81& 0.03& 20020109& UFTI\nl
\nl
SDSS J0354$-$0027&  17.14& 0.03& 16.46& 0.03& 15.95& 0.03& 20030104& UIST\nl
SDSS J0401$-$0609&  17.38& 0.03& 16.39& 0.03& 15.71& 0.03& 20030129& UFTI\nl
SDSS J0740$+$2009&  16.67& 0.03& 15.82& 0.03& 15.11& 0.03& 20030104& UIST\nl
SDSS J0741$+$2351&  15.87& 0.03& 16.12& 0.05& 16.12& 0.05& 20020217& UFTI\nl
SDSS J0742$+$2055&  15.60& 0.03& 15.95& 0.03& 16.06& 0.03& 20030104& UIST\nl
\nl
SDSS J0747$+$2937&  17.87& 0.05& 17.28& 0.05& 16.93& 0.05& 20020108& UFTI\nl
SDSS J0755$+$2934&  16.71& 0.03& 15.94& 0.03& 15.32& 0.03& 20020109& UFTI\nl
SDSS J0756$+$2314&  16.80& 0.03& 15.82& 0.03& 15.00& 0.03& 20030104& UIST\nl
SDSS J0758$+$3247&  14.78& 0.03& 14.21& 0.03& 13.87& 0.03& 20020217& UFTI\nl
SDSS J0805$+$4812&  14.61& 0.03& 14.01& 0.03& 13.51& 0.03& 20020109& UFTI\nl
\nl
SDSS J0809$+$4434&  16.37& 0.03& 15.25& 0.03& 14.31& 0.03& 20020109& UFTI\nl
SDSS J0830$+$0128&  15.99& 0.03& 16.17& 0.03& 16.38& 0.05& 20020108& UFTI\nl
SDSS J0831$+$3044&  17.45& 0.05& 16.89& 0.05& 16.35& 0.05& 20030105& UIST\nl
SDSS J0852$+$4720&  16.13& 0.03& 15.21& 0.03& 14.62& 0.03& 20020108& UFTI\nl
SDSS J0931$+$0327&  16.60& 0.05& 16.09& 0.05& 15.53& 0.05& 20020217& UFTI\nl
\nl
SDSS J0931$+$0327&  16.62& 0.03& 16.11& 0.03& 15.63& 0.03& 20030104& UIST\nl
SDSS J1004$+$0053&  17.40& 0.05& 16.82& 0.05& 16.24& 0.05& 20020620& UFTI\nl
SDSS J1030$+$0213&  17.10& 0.05& 16.27& 0.05& 15.67& 0.05& 20020108& UFTI\nl
SDSS J1044$+$0429&  15.84& 0.03& 14.97& 0.03& 14.32& 0.03& 20020109& UFTI\nl
SDSS J1046$+$0424&  16.97& 0.03& 16.03& 0.03& 15.35& 0.03& 20020620& UFTI\nl
\nl
SDSS J1104$+$5548&  17.31& 0.05& 16.71& 0.05& 16.31& 0.05& 20020109& UFTI\nl
SDSS J1104$+$5548&  17.26& 0.05& 16.75& 0.03& 16.38& 0.09& 20021204& UIST\nl
SDSS J1126$+$0120&  16.68& 0.03& 15.81& 0.03& 15.04& 0.03& 20020620& UFTI\nl
SDSS J1155$+$0559&  15.63& 0.03& 14.74& 0.03& 14.09& 0.03& 20020109& UFTI\nl
SDSS J1157$+$0611&  17.09& 0.05& 16.45& 0.05& 16.00& 0.05& 20010530& UFTI\nl
\nl
SDSS J1207$+$0244&  15.38& 0.03& 14.63& 0.03& 14.16& 0.03& 20020108& UFTI\nl
SDSS J1210$+$0307&  17.27& 0.05& 16.58& 0.03& 16.03& 0.03& 20020620& UFTI\nl
SDSS J1231$+$0847&  15.14& 0.03& 15.40& 0.03& 15.46& 0.03& 20020620& UFTI\nl
SDSS J1314$-$0008&  16.54& 0.03& 15.86& 0.03& 15.32& 0.03& 20020109& UFTI\nl
SDSS J1326$-$0038&  16.22& 0.03& 15.11& 0.03& 14.16& 0.03& 20030129& UFTI\nl
\nl
SDSS J1331$-$0116&  15.34& 0.03& 14.67& 0.03& 14.09& 0.03& 20020109& UFTI\nl
SDSS J1331$-$0116&  15.30& 0.03& 14.63& 0.03& 14.04& 0.03& 20020620& UFTI\nl
SDSS J1359$+$4728&  16.95& 0.03& 16.00& 0.03& 15.34& 0.03& 20020620& UFTI\nl
SDSS J1408$+$0539&  16.49& 0.03& 15.95& 0.03& 15.49& 0.03& 20030104& UIST\nl
SDSS J1432$-$0059&  16.99& 0.05& 16.14& 0.05& 15.45& 0.05& 20010530& UFTI\nl
\nl
SDSS J1435$-$0043&  16.41& 0.03& 15.68& 0.03& 15.12& 0.03& 20020620& UFTI\nl
SDSS J1440$+$0026&  15.93& 0.03& 15.23& 0.03& 14.64& 0.03& 20020620& UFTI\nl
SDSS J1516$+$0259&  16.88& 0.05& 16.07& 0.05& 15.35& 0.05& 20010530& UFTI\nl
SDSS J1521$+$0131&  16.06& 0.03& 15.63& 0.03& 15.48& 0.03& 20020620& UFTI\nl
SDSS J1525$+$5810&  16.90& 0.05& 16.13& 0.05& 15.43& 0.05& 20020109& UFTI\nl
\nl
SDSS J1616$+$2218&  17.53& 0.03& 16.50& 0.03& 15.65& 0.03& 20030618& UIST\nl
SDSS J1624$+$4441&  17.56& 0.05& 16.88& 0.05& 16.46& 0.05& 20010720& UFTI\nl
SDSS J1630$+$4344&  16.48& 0.03& 15.51& 0.03& 14.70& 0.03& 20020620& UFTI\nl
SDSS J1632$+$4150&  16.87& 0.05& 16.42& 0.08& 16.19& 0.08& 20020620& UFTI\nl
SDSS J1649$+$3842&  17.78& 0.10& 17.35& 0.10& 16.71& 0.10& 20010720& UFTI\nl
\nl
SDSS J1649$+$3842&  17.79& 0.10& 17.26& 0.10& 16.78& 0.10& 20020620& UFTI\nl
SDSS J1750$+$4222&  16.12& 0.03& 15.57& 0.03& 15.31& 0.03& 20020620& UFTI\nl
SDSS J1758$+$4633&  15.86& 0.03& 16.20& 0.03& 16.12& 0.03& 20020620& UFTI\nl
SDSS J2047$-$0718&  16.70& 0.03& 15.88& 0.03& 15.34& 0.03& 20011125& IRCAM\nl
SDSS J2124$+$0100 & 15.88& 0.03& 16.12& 0.03& 16.07& 0.03& 20031208& UFTI \nl
\nl
SDSS J2249$+$0044&  16.29& 0.03& 15.30& 0.03& 14.38& 0.03& 20020108& UFTI\nl

\enddata

\end{deluxetable}

\newpage

\begin{deluxetable}{lcccc}
\small
\tablenum{4}
\tablewidth{300pt}
\tablecaption{Spectroscopic Configurations}
\tablehead{
\colhead{Instrument}&  \colhead{Wavelength Range}&  \colhead{Resolution}& 
\colhead{Slit Width}\nl 
\colhead{Configuration}& \colhead{$\mu$m}& \colhead{\AA}& \colhead{arcsec}\nl
}
\tablecolumns{4}
\startdata 
CGS4--Z  & 0.84---1.04  & 25  & 1.20  \nl 
CGS4--J  & 1.03---1.35  & 21  & 1.20  \nl 
CGS4--H  & 1.38---2.02  & 50  & 1.20  \nl 
CGS4--K  & 1.88---2.52  & 50  & 1.20  \nl 
UIST--HK & 1.44---2.49  & 44  & 0.48  \nl 

\enddata

\end{deluxetable}

\newpage

\begin{deluxetable}{lccccc}
\small
\tablenum{5}
%\tablewidth{400pt}
\tablecaption{New  Spectroscopy}
\tablehead{
\colhead{Name}& \multicolumn{5}{c}{Date for Observation with Configuration} \nl 
\colhead{}&  \colhead{CGS4--Z}&  \colhead{CGS4--J}&  \colhead{CGS4--H}&
\colhead{CGS4--K}&  \colhead{UIST--HK}\nl
}
\tablecolumns{6}
\startdata 
2MASS J0415$-$0935& 20021214 & 20021213 &  20020110  & 20020111  & \nodata\nl 
2MASS J0727$+$1710& \nodata  & 20021212 &  \nodata  &  \nodata  &  20021212\nl 
2MASS J0929$+$3429& \nodata  &  \nodata  &  20011029  & \nodata &  \nodata\nl 
2MASS J0908$+$5032 & 20031208  &  20021212 & 20020110  & 20020111 & \nodata \nl
2MASS J0937$+$2931& \nodata  & 20020112 &  20020110  & 20020111 &  \nodata\nl 
\nl
2MASS J1439$+$1929& \nodata  & 20020625 &  20020619  & 20020622  & \nodata\nl 
2MASS J1507$-$1627& \nodata  & 20020625 &  20020624  & 20020624  & \nodata\nl 
2MASS J2224$-$0158& \nodata & 20020714 &  20020618  & 20020622 & \nodata\nl 
2MASS J2244$+$2043& \nodata& 20021214&  20020618,21& 20020618,22,23& \nodata\nl 
2MASS J2254$+$3123& 20030812& 20020714  & 20020109 & 20020112 & \nodata\nl 
\nl
2MASS J2339$+$1352&  \nodata & 20021214  & 20020625 & 20020625 & \nodata\nl 
SDSS J0000$+$2554 & \nodata & 20031208 & 20031207 & 20031207 & \nodata \nl
SDSS J0001$+$1535 & \nodata & \nodata &\nodata & \nodata & 20031120 \nl 
SDSS J0016$-$0043 & \nodata & \nodata &\nodata & \nodata &  20031120 \nl
SDSS J0127$+$1354& \nodata &  \nodata & 20020112  & 20020112 & \nodata\nl 
\nl
SDSS J0203$-$0108& \nodata   &   \nodata & \nodata & \nodata &  20030923 \nl
SDSS J0207$+$1355& \nodata &  \nodata &  20020111  &20020112 & 20030105 \nl
SDSS J0354$-$0027& \nodata   &   \nodata & \nodata & \nodata & 20030105 \nl
SDSS J0401$-$0609&\nodata   &   \nodata & \nodata & \nodata & 20030923 \nl
SDSS J0741$+$2351 & \nodata  &  20030924  & \nodata & \nodata & 20021130 \nl
\nl
SDSS J0742$+$2055 & \nodata  &  20030925 &  20030228  & 20030228  & \nodata\nl 
SDSS J0747$+$2937 & \nodata  &   \nodata & \nodata & \nodata & 20021214 \nl
SDSS J0755$+$2934 & \nodata  &   \nodata &  20020110 & 20020111 & 20021212 \nl
SDSS J0758$+$3247 & 20031208  &   20020225  & 20020225 &  20020225 & \nodata\nl 
SDSS J0805$+$4812 & 20031207  &  20031207 & \nodata &  \nodata &  20030321 \nl 
\nl
SDSS J0830$+$0128 & \nodata  &   \nodata &  20020110 & \nodata & \nodata \nl
SDSS J0852$+$4720 & \nodata  &  20031124 &  20020110 & 20020111 & \nodata \nl
SDSS J0931$+$0327 & \nodata  &   20031207 & \nodata &  \nodata & 20021212  \nl
SDSS J1004$+$0053 & \nodata  &   \nodata & \nodata &  \nodata & 20021214 \nl
SDSS J1030$+$0213 & \nodata  &   20031207 & 20020110  & 20020111 & \nodata \nl
\nl
SDSS J1044$+$0429 & \nodata  &   \nodata &  20020111  & 20020111 & \nodata \nl
SDSS J1046$+$0424 & \nodata  &   \nodata &   \nodata & 20020624 & \nodata \nl
SDSS J1104$+$5548 & \nodata   &   \nodata & \nodata & \nodata & 20021213 \nl
SDSS J1110$+$0116 & \nodata  &   \nodata & \nodata &  20020624 & \nodata \nl
SDSS J1126$+$0120 & \nodata  &  \nodata &  \nodata & 20020622 & \nodata \nl
\nl
SDSS J1155$+$0559& \nodata   &  \nodata &  20020111 & 20020112 & \nodata \nl
SDSS J1157$+$0611& \nodata   &   20031208 &  20020110 & 20020112 & \nodata \nl
SDSS J1207$+$0244 & 20040110  &  20020714    &  20020112 & 20020112 & \nodata \nl
SDSS J1210$+$0307 & \nodata  &  \nodata &  \nodata &   \nodata &   20021213 \nl
SDSS J1231$+$0847 & \nodata  &   20020625 & 20020619&  20020622 & \nodata \nl
\nl
SDSS J1314$-$0008 & \nodata  &     \nodata & \nodata & 20020623 & \nodata \nl
SDSS J1331$-$0116 & \nodata  &  20031207 &  20020112 & 20020622 & 20021213 \nl
SDSS J1359$+$4728 & \nodata   &  \nodata & 20020625 & 20020623 & \nodata \nl
SDSS J1432$-$0059 & \nodata   &  \nodata & 20020621 & 20020622 & \nodata \nl
SDSS J1435$-$0043 & \nodata   &  \nodata &  20020624 & \nodata & \nodata \nl
\nl
SDSS J1440$+$0026 & \nodata   &  \nodata & 20020624  & \nodata & \nodata \nl
SDSS J1516$+$0259 & \nodata   &   \nodata & 20020619 & 20020622 & \nodata \nl
SDSS J1521$+$0131 & \nodata   & 20030809& 20020619 & 20020625 & \nodata \nl
SDSS J1525$+$5810 & \nodata   &   \nodata & 20020621&  20020623,25 & \nodata \nl
SDSS J1616$+$2218 & \nodata  &  \nodata &  \nodata &   \nodata & 20030618  \nl
\nl
SDSS J1630$+$4344 & \nodata   &   \nodata & 20020624 & 20020622 & \nodata \nl
SDSS J1632$+$4150& \nodata& 20030821 & 20020624,20020714 & 20020625& \nodata \nl
SDSS J1750$+$4222 & \nodata& 20030812&   20020621& 20020622 & \nodata \nl
SDSS J1758$+$4633 & \nodata& 20030812&  20020618 & 20020622,23 & \nodata \nl
SDSS J2047$-$0718&  \nodata&   20030814&  20020718 & 20020807 & \nodata \nl
\nl
SDSS J2124$+$0100 & \nodata  & \nodata & 20031208 & \nodata & \nodata \nl
 
\enddata 

\end{deluxetable}

\newpage

\begin{deluxetable}{lrlrlrlrll}
\small
\tablenum{6}
%\tablewidth{400pt}
\tablecaption{Spectral Indices}
\tablehead{
\colhead{Name} & \multicolumn{2}{c}{H$_2$O(J)}& \multicolumn{2}{c}{H$_2$O(H)}& 
\multicolumn{2}{c}{CH$_4$(H)}& \multicolumn{2}{c}{CH$_4$(K)}& \colhead{Mean}\nl
\colhead{}& \colhead{Index}& \colhead{Type}& \colhead{Index} &  \colhead{Type}&
\colhead{Index}& \colhead{Type}& \colhead{Index}& \colhead{Type}&
\colhead{Type}\nl
}
\tablecolumns{10}
\startdata 
2MASS J0415$-$0935& 26.35& T9.5& 18.66& T9& 9.99 &  T8.5  & 25.26  & T9  &  T9\nl
2MASS J0727$+$1700& 10.69& T7.5& 11.69& T8.5& 6.09& T7.5& 16.26&  T8  &  T8\nl
2MASS J0908$+$5032& 1.41&[$<$T0] & 1.92& L9.5& 1.04& T0& 1.12& L7.5&  L9$\pm$1.0\nl
2MASS J0929$+$3429& \nodata& \nodata& 1.72& L7.5& 0.96& [$<$T0] & \nodata& \nodata& L7.5 \nl
2MASS J0937$+$2931\tablenotemark{a}& 6.05& T6.5& 6.85& T6.5& 3.70& T6.5& 4.39&  T5&  T6\nl
\nl
2MASS J1439$+$1929& 1.10& [$<$T0] & 1.30& L1& 0.96&[$<$T0] & 0.95&[$<$L3] & L1\nl
2MASS J1507$-$1627& 1.31& [$<$T0]& 1.59& L5.5& 0.98& [$<$T0]& 1.03& L5.5& L5.5\nl
2MASS J2224$-$0158& 1.37& [$<$T0]& 1.48& L3& 0.97& [$<$T0]& 0.95& L4& L3.5\nl
2MASS J2244$+$2043& 1.62& T0& 1.71& L7.5& 0.91& [$<$T0]& 0.99& L4.5&  L7.5$\pm$2.0\nl
2MASS J2254$+$3123& 2.46& T3.5& 3.69& T3.5& 1.83& T4.5& 3.36& T4&   T4\nl
\nl
2MASS J2339$+$1352& 4.08& T5& 7.26& T6.5& 2.35& T5.5& 5.03& T5& T5.5\nl
SDSS J0000$+$2554 & 3.03 & T4.5 & 4.11 & T4 & 1.84 & T4.5 &  3.70 & T4.5  & T4.5 \nl
SDSS J0001$+$1535 &  \nodata& \nodata& 1.59 & L5.5 & 0.95 & [$<$T0] &  0.92 & L3 & L4$\pm$1.0\nl
SDSS J0016$-$0043 & \nodata& \nodata& 1.61 & L5.5 & 0.85 &  [$<$T0] &  1.03 & L5.5 & L5.5\nl
SDSS J0127$+$1354& \nodata& \nodata& 1.58& L5& 0.96&[$<$T0] & 1.09& L7& L6$\pm$1.0\nl
\nl
SDSS J0203$-$0108& \nodata& \nodata&  2.00 & L9.5 & 1.04 & T0 &   1.21 & L9 & L9.5\nl
SDSS J0207$+$1355\tablenotemark{b}& \nodata& \nodata& 1.37& L2& 0.98&[$<$T0]  & 0.97& L4.5& L3$\pm$1.5\nl
SDSS J0354$-$0027& \nodata& \nodata& 1.31& L1& 1.00&[$<$T0] & 0.93& L3& L2$\pm$1.0\nl
SDSS J0401$-$0609& \nodata& \nodata&  1.43 & L2.5 & 0.98 & [$<$T0] & 1.01 & L5 & L4$\pm$1.5\nl
SDSS J0741$+$2351& 4.56 & T5.5 & 4.65& T4.5& 2.71& T5.5& 5.50& T5.5& T5.5\nl
\nl
SDSS J0742$+$2055& 3.88 & T5 & 4.78& T5& 2.74& T5.5& 5.10& T5.5&   T5\nl
SDSS J0747$+$2937& \nodata& \nodata& 1.07& M6& 0.94&[$<$T0] & 0.98& [L4.5]& late M\nl
SDSS J0755$+$2934\tablenotemark{b}& \nodata& \nodata& 1.41& L2.5& 0.99&[$<$T0] & 0.97& L4.5& L3.5$\pm$1.0\nl
SDSS J0758$+$3247& 2.47& T3.5& 2.67& T2& 1.12& T1& 1.59& T1.5& T2$\pm$1.0\nl
SDSS J0805$+$4812& 1.38 &  [$<$T0]  & 2.01& L9.5& 1.06& T0.5& 1.07& L6.5& L9$\pm$1.5\nl
\nl
SDSS J0830$+$0128& \nodata& \nodata& 6.25& T6& 2.44& T5.5& \nodata& \nodata& T5.5\nl
SDSS J0852$+$4720& 1.44 & [$<$T0]   & 2.14& T0.5& 1.07& T0.5& 1.15& L8&  L9.5$\pm$1.0\nl
SDSS J0931$+$0327&  1.46 & [$<$T0]& 1.89& L9& 0.90&[$<$T0] & 1.06& L6& L7.5$\pm$1.5\nl
SDSS J1004$+$0053& \nodata& \nodata& 1.29& L1& 0.92&[$<$T0] & 0.92& L3&  L2$\pm$1.0\nl
SDSS J1030$+$0213& 1.48 & [$<$T0] & 2.24& T1& 1.06& T0.5& 1.14& L8&  L9.5$\pm$1.0\nl
\nl
SDSS J1044$+$0429& \nodata& \nodata& 1.64& L6.5& 0.98& [$<$T0]& 1.11& L7& L7\nl
SDSS J1046$+$0424& \nodata& \nodata& \nodata& \nodata& \nodata& \nodata& 1.06& L6& L6\nl
SDSS J1104$+$5548& \nodata& \nodata& 1.98& L9.5&  1.09& T1& 1.17&  L7.5&  L9.5$\pm$1.5\nl
SDSS J1110$+$0116& 5.07& T6& 5.37& T5.5 & 2.47 & T5.5 & 6.28 & T6 & T5.5\nl
SDSS J1126$+$0120& \nodata& \nodata& \nodata& \nodata& \nodata& \nodata&  1.05&  L6& L6\nl
\nl
SDSS J1155$+$0559& \nodata& \nodata& 1.74& L8& 0.97&[$<$T0] & 1.09&  L7& L7.5\nl
SDSS J1157$+$0611& 1.78 & T1 &  3.00&  T2.5& 1.14&  T1.5&  1.45& T1&  T1.5\nl
SDSS J1207$+$0244& 1.64& T0& 2.15& T0.5& 1.06& T0.5& 1.24& L9.5& T0\nl
SDSS J1210$+$0307& \nodata& \nodata& 1.28& L0.5& 0.94& [$<$T0]& 0.91& L2.5& L1.5$\pm$1.0\nl
SDSS J1231$+$0847& 5.54& T6& 6.51& T6& 2.94& T6& 5.21& T5.5&  T6\nl
\nl
SDSS J1314$-$0008& \nodata& \nodata& 1.39& L2& 1.00&[$<$T0] & 0.98& L4.5& L3.5$\pm$1.5\nl
SDSS J1331$-$0116& 1.40 &[$<$T0] & 2.12,2.19& T0,T0.5& 1.00,0.96& [$<$T0]&   1.03,1.05&  L5.5,L6& L8$\pm$2.5\nl
SDSS J1359$+$4727& \nodata& \nodata& 1.77& L8& 1.00&[$<$T0] & 1.18& L8.5& L8.5\nl
SDSS J1432$-$0059& \nodata& \nodata& 1.53& L4& 0.95& [$<$T0]& 1.01& L5& L4.5\nl
SDSS J1435$-$0043& \nodata& \nodata& 1.43& L2.5& 0.97& [$<$T0]& \nodata&   \nodata& L2.5 \nl
\nl
SDSS J1440$+$0026& \nodata& \nodata& 1.32& L1& 0.98& [$<$T0]& \nodata&   \nodata& L1\nl
SDSS J1516$+$0259& \nodata& \nodata& 2.11& T0& 1.14& T1.5& 1.16& L8& T0$\pm$1.5\nl
SDSS J1521$+$0131& 2.07& T2& 3.07& T2.5& 1.10& T1& 2.01& T2.5&  T2\nl
SDSS J1525$+$5810& \nodata& \nodata& 1.70& L7.5& \nodata& \nodata& 1.03& L5.5&  L6.5$\pm$1.0\nl
SDSS J1616$+$2218& \nodata& \nodata& 1.46& L3& 0.95&[$<$T0] & 1.08& L6.5& L5$\pm$2.0\nl
\nl
SDSS J1630$+$4344& \nodata& \nodata& 1.76& L8& 1.01&[$<$T0] & 1.03& L5.5& L7$\pm$1.5\nl
SDSS J1632$+$4150& 1.59& T0& 2.13& T0& 1.15& T1.5& 1.81& T2& T1$\pm$1.0\nl
SDSS J1750$+$4222& 1.75& T1& 2.21& T0.5& 1.09& T1& 1.69& T1.5& T1 \nl
SDSS J1758$+$4633& 8.63& T7& 9.08& T7.5& 4.16& T6.5& 10.23& T7& T7\nl
SDSS J2047$-$0718& 1.42& [$<$T0]& 1.85& L9& 1.08& T0.5& 1.18& L8.5&  L9.5$\pm$1.0\nl
\nl
SDSS J2124$+$0100 & \nodata  & \nodata & 6.42 &  T6 & 2.40 & T5.5 & \nodata & \nodata & T6 \nl

\nl
\tablenotetext{a}{Mean type includes the $K$-band index for this dwarf with 
suppressed $K$-band flux, see \S 5.6.}
\tablenotetext{b}{ Based on UIST data only, more consistent than 
previous CGS4 data.}

\enddata 

\end{deluxetable}

\newpage

\begin{deluxetable}{lcccc}
\small
\tablenum{7}
\tablewidth{400pt}
\tablecaption{Provisional Spectral Indices for the End of the T Sequence}
\tablehead{
\colhead{Type} &  \colhead{H$_2$O 1.2$\mu$m} &  \colhead{H$_2$O 1.5$\mu$m}  &
\colhead{CH$_4$ 1.6$\mu$m } &  \colhead{CH$_4$ 2.2$\mu$m} \nl 
}
\tablecolumns{5}
\startdata 
T8  &  10---15      &             9---12     &     6---9       &      12---18\nl 
T9   &  15---30     &             12---24    &      9---18     &       18---36\nl 
T10  &  $>$30       &               $>$24    &       $>$18     &         $>$36\nl 

\enddata

\end{deluxetable}

\newpage
\begin{deluxetable}{lrrrrll}
\tablenum{8}
%\tablewidth{250pt}
\tablecaption{Parallaxes and Absolute K Magnitudes}
\tablehead{
\colhead{Name} &  \colhead{$\pi$\tablenotemark{a}}&  \colhead{error}& 
\colhead{$\rm M_K$}& \colhead{error}& \colhead{Spectral}& 
\colhead{Reference\tablenotemark{b}}\nl
\colhead{} &  \colhead{mas}&  \colhead{$\pi$ mas}& 
\colhead{}& \colhead{$\rm M_K$}& \colhead{Type}& 
\colhead{}\nl
}
\tablecolumns{7}
\startdata

2MASS J0030$-$1450&       37.4&   4.5& 12.36& 0.26& L7&   V03\nl 
2MASS J0036$+$1821&      114.2&   0.8& 11.33& 0.03& L4&   D02\nl 
2MASS J0328$+$2302&       33.1&   4.2& 12.47& 0.28& L9.5& V03\nl
2MASS J0345$+$2540&       37.1&   0.5& 10.51& 0.04& L1&   D02\nl 
2MASS J0415$-$0935&      174.3&   2.8& 17.04& 0.04& T9&   V03\nl
\nl
2MASS J0559$-$0559&       96.7&   1.0& 13.66& 0.04& T4.5& D02,V03\nl
2MASS J0727$+$1710&      110.1&   2.3& 15.90& 0.05& T8&   V03\nl
2MASS J0746$+$2000AB&     81.9&   0.3& 10.00& 0.03& L1&   D02\nl 
2MASS J0825$+$2115&       94.2&   0.9& 12.80& 0.04& L6&   D02,V03\nl
2MASS J0850$+$1057AB&     33.8&   2.7& 12.00& 0.18& L6&   D02,V03\nl
\nl
2MASS J0937$+$2931&      162.8&   3.9& 16.45& 0.08& T6& V03\nl
2MASS J1047$+$2124&       98.8&   3.3& 16.17& 0.08& T6.5& V03,T03\nl
2MASS J1217$-$0311&       93.2&   2.1& 15.77& 0.06& T8&   V03,T03\nl 
2MASS J1225$-$2739AB&     74.8&   2.0& 14.65& 0.07& T6&   V03,T03\nl 
2MASS J1439$+$1929&       69.6&   0.5& 10.68& 0.03& L1&   D02\nl
\nl
2MASS J1507$-$1627&      136.4&   0.6& 11.96& 0.03& L5.5&   D02\nl
2MASS J1523$+$3014\tablenotemark{c}&       54.4&   1.1& 13.03& 0.06& L8&   Hip,YPC,V03\nl
2MASS J1534$-$2952AB&     73.6&   1.2& 14.24& 0.06& T5.5& T03\nl
2MASS J1632$+$1904&       65.0&   1.8& 13.03& 0.07& L7.5& D02,V03\nl
2MASS J2224$-$0158&       87.0&   0.9& 11.68& 0.04& L3.5& D02,V03\nl
\nl
2MASS J2356$-$1553&       69.0&   3.4& 14.92& 0.11& T6&   V03\nl  
DENIS-P~J0205.4$-$1159AB&  50.6&   1.5& 11.51& 0.07& L5.5& D02\nl
DENIS-P~J1058.7$-$1548&    57.7&   1.0& 11.36& 0.05& L3&   D02\nl  
DENIS-P~J1228.2$-$15AB&    49.4&   1.9& 11.18& 0.09& L6&   D02\nl  
GD 165B&                  31.7&   2.5& 11.59& 0.17& L3&   YPC\nl  
\nl
Gl 229B&                 173.2&   1.1& 15.55& 0.03& T6&   Hip,YPC\nl
Gl 570D&                 170.2&   1.4& 16.67& 0.04& T8&   Hip,YPC\nl
Kelu-1&                  53.6&   2.0& 10.43& 0.09& L3&   D02\nl     
LHS 102B&                104.7&  11.4& 11.46& 0.24& L4.5& YPC\nl 
SDSS J0032$+$1410&        30.1&   5.2& 12.39& 0.37& L8&   V03\nl 
\nl
SDSS J0107$+$0041&        64.1&   4.5& 12.61& 0.16& L5.5& V03\nl 
SDSS J0151$+$1244&        46.7&   3.4& 13.53& 0.16& T1&   V03\nl  
SDSS J0207$+$0000&        34.9&   9.9& 14.33& 0.62& T4.5& V03\nl
SDSS J0423$-$0414&        65.9&   1.7& 12.05& 0.06& T0&   V03\nl
SDSS J0539$-$0059&        76.1&   2.2& 11.81& 0.07& L5&   V03\nl
\nl
SDSS J0830$+$4828&        76.4&   3.4& 13.10& 0.10& L9&   V03\nl
SDSS J0837$-$0000&        33.7&  13.5& 13.62& 0.87& T0.5& V03\nl
SDSS J1021$-$0304&        35.4&   4.2& 13.00& 0.26& T3&   V03,T03\nl
SDSS J1254$-$0122&        74.0&   1.6& 13.22& 0.06& T2&   V03,T03\nl
SDSS J1326$-$0038&        50.0&   6.3& 12.66& 0.28& L5.5& V03\nl
\nl
SDSS J1346$-$0031&        69.1&   2.1& 14.93& 0.07& T6&   V03,T03\nl
SDSS J1435$-$0043&        16.1&   5.8& 11.15& 0.78& L2.5& V03\nl
SDSS J1446$+$0024&        45.5&   3.2& 12.09& 0.16& L5&   V03\nl
SDSS J1624$+$0029&        90.7&   1.0& 15.40& 0.04& T6&   D02,V03,T03\nl
SDSS J1750$+$1759&        36.2&   4.5& 13.82& 0.27& T3.5& V03\nl

\tablenotetext{a}{Parallaxes are weighted means of the values given in the 
cited references (see Golimowski et al. 2004).}
\tablenotetext{b}{References for parallax: \nl
Hip: Hipparcos Astrometric Catalogue: ESA (1997), Perryman et al. (1997)\nl
YPC: Yale Parallax Catalogue (van Altena et al. 1995)\nl
D02: Dahn et al. (2002)\nl
V03: Vrba et al. (2004)\nl
T03: Tinney et al. (2003)}
\tablenotetext{c}{2MASS J1523+3014 = Gl 584C.}
\enddata

\end{deluxetable}

\newpage

\begin{deluxetable}{lllrrrrrrrrr}
%\small
\scriptsize
\tablenum{9}
%\tablewidth{400pt}
\tablecaption{Magnitudes and Colors, Sorted by Infrared Spectral Type}
\tablehead{
\colhead{Name}& \colhead{Type\tablenotemark{a}}& \colhead{$M_J$}&  \colhead{$i$--$z$}& 
\colhead{$z$}& \colhead{$z$--$J$} & \colhead{$Z$--$J$} & \colhead{$J$}& 
\colhead{$J$--$K$} &\colhead{$J$--$H$} &\colhead{$H$--$K$}\nl 
}
\tablecolumns{12}
\startdata
SDSS J083120.81$+$304417.1& [M--L2]& \nodata& 1.77& 19.68& 2.23& \nodata& 17.45& 1.10& 0.56& 0.54\nl
SDSS J140814.74$+$053952.9& [M--L2]& \nodata& 1.77& 18.71& 2.22& \nodata& 16.49& 1.00& 0.54& 0.46\nl
SDSS J162441.00$+$444145.8& [M--L3]&\nodata& \nodata& 20.07& 2.51& \nodata& 17.56& 1.10& 0.68& 0.42\nl
%SDSSJ074719.71$+$293748.6& L0$\pm$4& \nodata& \nodata& 19.91& 2.04& \nodata& %17.87& 0.94& 0.59& 0.35\nl
2MASS J03454316$+$2540233\tablenotemark{b}& L1$\pm$1&  11.69&  \nodata&\nodata& \nodata& \nodata& 13.84& 1.18& 0.63& 0.54\nl  
2MASS J07464256$+$2000321AB& L1&  11.21& 1.80& 14.30& 2.66& 1.47& 11.64&1.21& 0.63& 0.58\nl
\nl
2MASS J14392836$+$1929149 & L1\tablenotemark{c}&  11.87& \nodata&\nodata& \nodata& \nodata& 12.66& 1.19& 0.61& 0.58\nl  
SDSS J144016.20$+$002638.9& L1\tablenotemark{d} & \nodata& 1.90& 18.75& 2.82& \nodata& 15.93& 1.29& 0.70& 0.59\nl
SDSS J121001.96$+$030739.2& L1.5$\pm$1 & \nodata& \nodata& 19.79& 2.52& \nodata& 17.27& 1.24& 0.69& 0.55\nl
SDSS J035448.73$-$002742.1& L2$\pm$1 & \nodata& 2.02& 19.61& 2.47& \nodata&17.14& 1.19& 0.68& 0.51\nl
SDSS J100401.41$+$005354.9& L2$\pm$1& \nodata& 2.05& 19.76& 2.36& \nodata& 17.40& 1.16& 0.58& 0.58\nl
\nl
SDSS J143535.70$-$004347.0& L2.5& 12.44& 1.84& 19.02& 2.61& \nodata& 16.41&  1.29&  0.73&  0.56\nl
2MASS J00283943$+$1501418\tablenotemark{e}&L3\tablenotemark{d}& \nodata& 2.19& 19.52& 3.03& 2.01& 16.49&  1.95&  1.01&  0.94\nl
DENIS-P~J1058.7$-$1548\tablenotemark{b,f}&  L3& 12.93&  \nodata& \nodata& \nodata& 1.64& 14.12& 1.57& 0.84& 0.74\nl 
GD 165B\tablenotemark{b,g}& L3$\pm$2 &13.14& \nodata& \nodata& \nodata& \nodata&  15.64&  1.55& 0.89& 0.66\nl 
Kelu--1\tablenotemark{b,h}& L3$\pm$1&11.88& \nodata& \nodata& \nodata&1.77&  13.23& 1.45& 0.78& 0.67\nl
\nl
SDSS J020735.60$+$135556.3& L3$\pm$1.5 & \nodata&  1.78& 18.06& 2.79& \nodata& 15.27& 1.46&  0.82&  0.64\nl
2MASS J22244381$-$0158521& L3.5& 13.59& \nodata& \nodata& \nodata& \nodata& 13.89& 1.91& 1.05& 0.86\nl 
SDSS J075515.26$+$293445.4& L3.5$\pm$1&\nodata& 2.33& 19.40& 2.69& \nodata& 16.71& 1.39& 0.77&  0.62\nl
SDSS J131415.52$-$000848.1& L3.5$\pm$1.5&\nodata&2.00& 19.63& 3.11& \nodata& 16.52& 1.22&0.68& 0.55\nl   
2MASS J00361617$+$1821104& L4$\pm$1&12.61&\nodata& \nodata& \nodata& 1.81& 12.30& 1.26&0.66& 0.60\nl 
\nl
SDSS J000112.18$+$153535.5 &  L4$\pm$1& \nodata& 1.74 & 18.55 & 3.26&\nodata& 15.29 & 1.77 & 0.89 & 0.88 \nl 
SDSS J040100.96$-$060933.0& L4$\pm$1.5 &\nodata& \nodata& 20.19& 2.81& \nodata&17.38& 1.67& 0.99& 0.68\nl 
LHS 102B\tablenotemark{i}& L4.5& 13.16& \nodata& \nodata& \nodata&  1.61& 13.06& 1.70&0.92& 0.78\nl 
SDSS J143211.74$-$005900.8& L4.5& \nodata&\nodata& 19.62& 2.63& \nodata& 16.99& 1.54& 0.85& 0.69\nl 
SDSS J053951.99$-$005902.0& L5& 13.26& 2.24& 16.78& 2.93& 1.75& 13.85& 1.45&0.81& 0.64\nl 
\nl 
SDSS J125737.26$-$011336.1& L5 & 13.86&2.14& 18.56& 2.92& \nodata& 15.64& 1.58& 0.96&0.62\nl
SDSS J144600.60$+$002452.0& L5&  13.85& 2.30& 18.48& 2.92& \nodata& 15.56&1.76& 0.97& 0.79\nl 
SDSS J161626.46$+$221859.2& L5$\pm$2& \nodata& \nodata& 20.33& 2.80& \nodata&17.53& 1.88& 1.03& 0.85\nl 
SDSS J224953.45$+$004404.2\tablenotemark{j}&L5$\pm$1.5\tablenotemark{k}&  \nodata& 2.16& 19.42& 2.95 & 1.77& 16.47&2.05& 1.11& 0.94\nl  
SDSS J074007.30$+$200921.9& [L3--8]& \nodata& \nodata&  19.78& 3.11& \nodata&16.67& 1.56&  0.85&  0.71\nl
\nl
SDSS J075656.54$+$231458.5&[L3--8]& \nodata& \nodata&  19.82& 3.02& \nodata& 16.80& 1.80& 0.98&  0.82\nl
2MASS J15074769$-$1627386& L5.5& 13.37& \nodata& \nodata& \nodata& \nodata& 12.70& 1.41&0.80& 0.61\nl
DENIS-P~J0205.4$-$1159AB\tablenotemark{l}&  L5.5$\pm$2\tablenotemark{d}& 12.95& \nodata& \nodata& \nodata& 1.58&  14.43& 1.44&  0.82& 0.62\nl 
SDSS J001608.44$-$004302.3 &  L5.5& \nodata& 1.77 & 19.34 & 3.00 &  \nodata& 16.34 & 1.82 & 1.00 & 0.82 \nl
SDSS J010752.33$+$004156.1&  L5.5\tablenotemark{d}& 14.78& 2.83& 18.70& 2.95& 1.58&  15.75&2.17& 1.19& 0.98\nl 
\nl
SDSSJ132629.82$-$003831.5&  L5.5& 14.70& 2.63& 19.05& 2.84& \nodata& 16.21&2.04& 1.11& 0.93\nl
DENIS-P~J1228.2$-$1547AB\tablenotemark{b,m}& L6$\pm$2& 12.75& \nodata&\nodata& \nodata&  1.73&  14.28& 1.57& 0.88& 0.69\nl
2MASS J08251968$+$2115521& L6\tablenotemark{d}& 14.76& \nodata& \nodata& \nodata&  1.73& 14.89& 1.96& 1.08& 0.88\nl 
2MASS J08503593$+$1057156AB & L6\tablenotemark{c}&13.85& \nodata& \nodata& \nodata& 1.95& 16.20&1.85& 0.99& 0.86\nl 
SDSS J012743.50$+$135420.9& L6$\pm$1&\nodata&  2.65& 19.62& 2.91 & \nodata&16.71& 1.62&  0.87&  0.75\nl 
\nl
SDSS J104625.76$+$042441.0& L6& \nodata& 2.64&  19.74& 2.77& \nodata& 16.97& 1.62& 0.94& 0.68\nl 
SDSS J112615.25$+$012048.2& L6&  \nodata& \nodata& 19.79& 3.11& \nodata& 16.68&1.64& 0.87& 0.77\nl 
SDSS J080959.01$+$443422.2& [L5--8]&  \nodata& 2.54& 19.28& 2.91& \nodata& 16.37& 2.06& 1.12& 0.94\nl 
2MASS J08014056$+$4628498 &  L6.5\tablenotemark{c} &\nodata  & 2.50& 18.78& 2.57& \nodata&16.21& 1.63& 0.90& 0.73\nl
SDSS J023617.93$+$004855.0& L6.5& \nodata& 2.87& 18.80& 2.79 & 1.53& 16.01&1.47& 0.85& 0.62\nl
\nl
SDSS J152531.32$+$581053.1& L6.5$\pm$1&  \nodata& \nodata& 19.75& 2.85& \nodata& 16.90& 1.47& 0.77& 0.70\nl
2MASS J00303013$-$1450333& L7\tablenotemark{c}& 14.26& \nodata& \nodata& \nodata& 1.74& 16.39& 1.90& 1.02& 0.88\nl
SDSS J104409.43$+$042937.6& L7&\nodata& 2.94& 18.73& 2.89& \nodata& 15.84& 1.52& 0.87& 0.65\nl
SDSS J163030.53$+$434404.0& L7$\pm$1.5&\nodata&  2.65& 19.45& 2.97& \nodata& 16.48& 1.78& 0.97& 0.81\nl
2MASS J09293364$+$3429527& L7.5& \nodata& \nodata& 19.52& 2.83& 1.69& 16.69&1.96& 1.07& 0.89\nl
\nl  
2MASS J16322911$+$1904407& L7.5&  14.83& 3.20& 18.58& 2.81& 1.75& 15.77&1.80& 1.09& 0.71\nl  
2MASS J22443167$+$2043433& L7.5$\pm$2& \nodata& \nodata& \nodata& \nodata& \nodata& 16.33 & 2.43& 1.27& 1.16\nl
SDSS J093109.56$+$032732.5& L7.5$\pm$1.5& \nodata& 2.72& 19.28& 2.68& \nodata& 16.60& 1.07& 0.51& 0.56\nl
SDSS J115553.86$+$055957.5& L7.5& \nodata& 2.81& 18.45& 2.82& \nodata& 15.63&1.54& 0.89& 0.65\nl
2MASS J15232263$+$3014562\tablenotemark{n}& L8& 14.63& \nodata& \nodata& \nodata& 1.65& 15.95&1.60& 0.90& 0.70\nl
\nl
SDSS J003259.36$+$141036.6& L8& 13.98&\nodata&  19.40& 2.82& 1.67& 16.58&1.59& 0.92& 0.67\nl
SDSS J085758.45$+$570851.4& L8$\pm$1& \nodata& 2.98& 17.77& 2.97& 1.72& 14.80& 1.86& 1.00& 0.86\nl 
SDSS J133148.90$-$011651.4& L8$\pm$2.5\tablenotemark{k} & \nodata&  2.42& 18.14& 2.82& \nodata& 15.32&1.25& 0.67& 0.58\nl
SDSS J135923.99$+$472843.2& L8.5& \nodata& \nodata& 19.76& 2.81& \nodata& 16.95&1.61& 0.95& 0.66\nl
2MASS J03105986$+$1648155& L9& \nodata& \nodata& \nodata& \nodata& \nodata&15.88& 1.70& 0.97& 0.73\nl 
\nl
2MASS J09083803$+$5032088& L9$\pm$1\tablenotemark{k}& \nodata& 2.84& 17.22& 2.82& \nodata& 14.40& 1.51& 0.86& 0.65\nl
SDSS J080531.80$+$481233.0& L9$\pm$1.5\tablenotemark{k}& \nodata& 2.20& 17.62& 3.01& \nodata& 14.61&1.10& 0.60& 0.50\nl
SDSS J083008.12$+$482847.4& L9$\pm$1& 14.64& 3.14& 18.08& 2.86& 1.64& 15.22& 1.54& 0.82& 0.72\nl 
2MASS J03284265$+$2302051& L9.5\tablenotemark{k}& 13.95& \nodata& \nodata& \nodata& 1.71& 16.35& 1.48& 0.88& 0.60\nl 
SDSS J020333.26$-$010812.5& L9.5 & \nodata& \nodata&  20.36& 2.53& \nodata&17.83& 1.65& 0.96& 0.69\nl
\nl
SDSS J085234.90$+$472035.0& L9.5$\pm$1& \nodata& 3.09& 18.90& 2.77& \nodata&16.13& 1.51& 0.92& 0.59\nl 
SDSS J103026.78$+$021306.4& L9.5$\pm$1& \nodata& \nodata& 19.94& 2.84& \nodata&17.10& 1.43& 0.83& 0.60\nl
SDSS J110454.25$+$554841.4& L9.5$\pm$1.5& \nodata& \nodata& 19.94& 2.66& \nodata& 17.28& 0.94& 0.55& 0.39\nl 
SDSS J204749.61$-$071818.3& L9.5$\pm$1& \nodata& \nodata& 19.74& 3.04& \nodata&16.70& 1.36& 0.82& 0.54\nl 
SDSS J042348.57$-$041403.5& T0\tablenotemark{k}& 13.39& 2.89& 17.29& 2.99& 1.68& 14.30& 1.34&0.79& 0.55\nl
\nl   
SDSS J120747.17$+$024424.8& T0\tablenotemark{k}& \nodata& 3.06& 18.41& 3.03& \nodata& 15.38& 1.22& 0.75& 0.47\nl
SDSS J151603.03$+$025928.9& T0$\pm$1.5& \nodata& \nodata& 19.89& 3.01& \nodata& 16.88& 1.53& 0.81& 0.72\nl
SDSS J083717.21$-$000018.0& T0.5& 14.54& \nodata& 20.06& 3.16& 1.69& 16.90&0.92& 0.69& 0.23\nl 
SDSS J015141.69$+$124429.6& T1$\pm$1& 14.60& \nodata& 19.45& 3.20& 1.84& 16.25&1.07& 0.71& 0.36\nl 
SDSS J163239.34$+$415004.3& T1$\pm$1& \nodata& \nodata& 20.35& 3.48& \nodata& 16.87& 0.68& 0.45& 0.23\nl
\nl 
SDSS J175024.01$+$422237.8& T1& \nodata& \nodata& 19.38& 3.26 & \nodata& 16.12&0.81& 0.55& 0.26\nl 
SDSS J115700.50$+$061105.2& T1.5 & \nodata& \nodata& 20.20& 3.11& \nodata& 17.09& 1.09& 0.64& 0.45\nl
SDSS J075840.33$+$324723.4& T2$\pm$1& \nodata& 3.96& 17.96& 3.18& \nodata& 14.78& 0.91& 0.57& 0.34\nl
SDSS J125453.90$-$012247.4& T2& 14.00& 4.22& 18.03& 3.37& 1.74& 14.66& 0.82&0.53& 0.29\nl 
SDSS J152103.24$+$013142.7& T2& \nodata& \nodata& 19.57& 3.51& \nodata& 16.06&0.58& 0.43& 0.15\nl
\nl
SDSS J102109.69$-$030420.1& T3& 13.62& \nodata& 19.33& 3.45& 1.78& 15.88& 0.62&0.47& 0.15\nl
SDSS J175032.96$+$175903.9& T3.5&13.94& \nodata& 19.63& 3.49& \nodata& 16.14&0.12& 0.20& --0.08\nl  
2MASS J22541892$+$3123498& T4& \nodata& \nodata& \nodata& \nodata& \nodata&15.01& --0.02& 0.06& --0.08\nl 
2MASS J05591914$-$1404488& T4.5& 13.50& \nodata& \nodata& \nodata& 1.98& 13.57&--0.16& --0.07& --0.09\nl
SDSS J000013.54$+$255418.6 & T4.5 &  \nodata& \nodata& 18.48 & 3.75 &  \nodata& 14.73 & --0.09 & --0.01 & --0.08\nl 
\nl
SDSS J020742.83$+$000056.2& T4.5&14.34& \nodata& 20.11& 3.48& 2.08& 16.63&0.01& --0.03& 0.04\nl  
SDSS J092615.38$+$584720.9& T4.5&\nodata& \nodata& 19.01& 3.54& \nodata& 15.47&--0.03&  0.05& --0.08\nl    
2MASS J07554795$+$2212169& T5\tablenotemark{o}&\nodata& \nodata& 19.12& 3.66& \nodata& 15.46& --0.40& --0.24& --0.16\nl
SDSS J074201.41$+$205520.5& T5& \nodata& \nodata& 19.28& 3.68& \nodata& 15.60&--0.46& --0.35& --0.11\nl
2MASS J15031961$+$2525196& T5.5\tablenotemark{o}& \nodata& \nodata& \nodata& \nodata& \nodata& 13.55& --0.44& --0.35& --0.09\nl
\nl
2MASS J15344984$-$2952274AB& T5.5\tablenotemark{o}& 13.93& \nodata& \nodata&\nodata& \nodata& 14.60&  --0.31&--0.14&--0.17\nl
2MASS J23391025$+$1352284& T5.5&\nodata& \nodata& \nodata& \nodata& \nodata&15.81& --0.36& --0.19& --0.17\nl
SDSS J074149.15$+$235127.5& T5.5&\nodata& \nodata& 19.65& 3.78& \nodata& 15.87&--0.35& --0.25& --0.10\nl
SDSS J083048.80$+$012831.1& T5.5& \nodata& \nodata& 19.59& 3.60& \nodata& 15.99&--0.39& --0.18& --0.21\nl 
SDSS J111010.01$+$011613.1& T5.5 & \nodata& \nodata& 19.64& 3.52& \nodata& 16.12&0.07& --0.10& 0.17\nl 
\nl
2MASS J02431371$-$2453298& T6\tablenotemark{o}& 14.99& \nodata& \nodata&\nodata& 1.85& 15.13& --0.21& --0.26& 0.05\nl
2MASS J09373487$+$2931409& T6 & 15.35& \nodata& \nodata& \nodata& 1.72& 14.29& --1.10& --0.38& --0.72\nl
2MASS J12255432$-$2739466AB& T6& 14.25& \nodata& \nodata& \nodata& 1.89& 14.88&--0.40& --0.29& --0.11\nl 
2MASS J23565477$-$1553111& T6\tablenotemark{o}& 14.67& \nodata& \nodata& \nodata& 2.12& 15.48& --0.25& --0.22&--0.03\nl
Gl 229B\tablenotemark{p}& T6$\pm$1&15.20& \nodata& \nodata& \nodata& 2.17& 14.01& --0.35&--0.35& 0.00\nl
\nl
SDSS J123147.39$+$084730.7& T6& \nodata& \nodata& 18.94& 3.80& \nodata& 15.14&--0.32& --0.26& --0.06\nl
SDSS J134646.45$-$003150.4\tablenotemark{b}& T6 &  14.69& \nodata& 19.29& 3.80&2.24& 15.49& --0.24& --0.35&  0.11\nl
SDSS J162414.37$+$002915.6\tablenotemark{b}& T6 &  14.99& \nodata& 19.02& 3.82&2.12& 15.20& --0.41& --0.28& --0.13\nl  
SDSS J212413.89$+$010000.3 & T6 &  \nodata& \nodata& 19.71 & 3.83 & \nodata& 15.88 & --0.19 & --0.24 & 0.05 \nl
2MASS J10475385$+$2124234& T6.5& 15.43& \nodata& \nodata& \nodata& 1.93&  15.46& --0.74& --0.37& --0.37\nl 
\nl
SDSS J175805.46$+$463311.9& T7 &  \nodata& \nodata& 19.67& 3.81& \nodata&15.86& --0.26& --0.34& 0.08\nl 
2MASS J15530228$+$1532369AB& T7.5\tablenotemark{o}&\ \nodata& \nodata& \nodata&\nodata& \nodata&15.34& --0.60& --0.42& --0.18\nl 
2MASS J07271824$+$1710012& T8&  15.40& \nodata& \nodata& \nodata& 1.98& 15.19&--0.50& --0.48& --0.02\nl 
2MASS J12171110$-$0311131& T8&  15.41& \nodata& 19.18& 3.62& 2.00& 15.56& --0.36& --0.42& 0.06\nl 
Gl 570D\tablenotemark{q}& T8&  15.97& \nodata& \nodata& \nodata& 1.92& 14.82&--0.70& --0.46& --0.24\nl   
\nl
2MASS J04151954$-$0935066& T9&  16.53& \nodata& \nodata& \nodata& 1.98& 15.32&--0.51& --0.38&--0.13\nl
\nl

\tablebreak
\tablenotetext{a}{Spectral types in square brackets estimated from colors.}
\tablenotetext{b}{$JHK$ magnitudes synthesized from flux calibrated spectrum.}
\tablenotetext{c}{Spectral type from Kirkpatrick et al. (2000).}
\tablenotetext{d}{Optical spectral type later than infrared type by $>$1 subclass.}
\tablenotetext{e}{2MASS J0028$+$1501 may be variable at the 5\% level; measured
$J/H/K$ of 16.52(0.03) / 15.42(0.03) / 14.63(0.05) on 20001119,
16.65(0.05) / 15.56(0.05) / 14.57(0.05) on 20010114,
16.45(0.03) / 15.54(0.03) / 14.51(0.03) on 20011124 and
16.44(0.03) / 15.44(0.03) / 14.53(0.03) on 20020108.}
\tablenotetext{f}{DENIS-P~J1058.7$-$1548 = 2MASS J10584787$-$1548172.}
\tablenotetext{g}{GD 165B = 2MASS J14243909$+$0917104.}
\tablenotetext{h}{Kelu-1 = 2MASS J13054019$-$2541059.}
\tablenotetext{i}{LHS 102B = 2MASS J00043484$-$4044058.}
\tablenotetext{j}{SDSS J2249$+$0044 may be variable at the 10\% level;
measured $J/H/K$ of
16.58(0.03) / 15.39(0.03) / 14.46(0.03)  on 20001119,
16.46(0.05) / 15.42(0.05) / 14.43(0.05) on 20010116,
16.36(0.03) / 15.30(0.03) / 14.38(0.03) on 20020108.}
\tablenotetext{k}{Optical spectral type earlier than infrared type by $>$1 subclass.}
\tablenotetext{l}{DENIS-P~J0205.4$-$1159AB =  2MASS J02052940$-$1159296AB.}
\tablenotetext{m}{DENIS-P~J1228.2$-$1547AB = 2MASS J12281523$-$1547342AB.}
\tablenotetext{n}{2MASS J15232263$+$3014562 = Gl 584C.}
\tablenotetext{o}{Spectral type from Burgasser et al. (2002)
or typed by us from Burgasser et al. data.}
\tablenotetext{p}{$ZJHK$ magnitudes synthesized from flux calibrated spectrum.}
\tablenotetext{q}{Gl 570D = 2MASS J14571496$-$2121477.}

\enddata

\end{deluxetable}

\newpage
\begin{deluxetable}{lrr}
\tablenum{10}
\tablewidth{250pt}
\tablecaption{SDSS $r$ Magnitudes for Detected Objects in Table 9}
\tablehead{
\colhead{Name} &  \colhead{$r$} & 
\colhead{$\sigma_r$}\nl
}
\tablecolumns{3}
\startdata
 
2MASS J0746$+$2000AB&  18.65& 0.01\nl
2MASS J0755$+$2212&  22.27& 0.23\nl
2MASS J0908$+$5032&   22.43& 0.17\nl 
SDSS J0001$+$1535 & 23.02 &  0.22 \nl
SDSS J0207$+$1355&   22.37& 0.14 \nl
SDSS J0423$-$0414&   22.82& 0.22 \nl
SDSS J0539$-$0059&   21.42& 0.08 \nl  
SDSS J0755$+$2934&  22.96& 0.29\nl  
SDSS J0805$+$4812&   22.79& 0.27\nl
SDSS J1257$-$0113&   22.68& 0.27\nl 
SDSS J1314$-$0008&   23.32& 0.27\nl
SDSS J1331$-$0116&   22.91& 0.26\nl 
SDSS J1440$+$0026&   22.92& 0.19\nl
SDSS J1446$+$0024&   23.20& 0.23\nl 

\tablenotetext{}{$r$ magnitude for zero flux = 25.1.}
\enddata

\end{deluxetable}
\newpage
\begin{deluxetable}{lll}
\tablenum{11}
\tablewidth{400pt}
\tablecaption{Surface Gravities of T5 and Later Dwarfs Estimated from $H-K$
Color}
\tablehead{
\colhead{log~$g\approx$4.5} & \colhead{log~$g\approx$5.0} & \colhead{log~$g\approx$5.5}\nl
}
\tablecolumns{3}
\startdata
2MASS J0243$-$2453 &  2MASS J0415$-$0935   & 2MASS J0755$+$2212\nl
2MASS J0727$+$1710 &  2MASS J1225$-$2739AB & 2MASS J0937$+$2931 \nl
2MASS J1217$-$0311 &  2MASS J1503$+$2525  & 2MASS J1047$+$2124  \nl
Gl 229B  & 2MASS J1553$+$1532AB  & 2MASS J1534$-$2952AB\nl
SDSS J1110$+$0116 &  2MASS J2356$-$1553  & 2MASS J2339$+$1352  \nl
SDSS J1346$-$0031  & Gl 570D              & SDSS J0830$+$0128 \nl
SDSS J1758$+$4633  &  SDSS J0741$+$2351  &  \nodata \nl 
SDSS J2124$+$0100  &SDSS J0742$+$2055  &\nodata \nl 
\nodata       &SDSS J1231$+$0847   &\nodata \nl 
\nodata       &SDSS J1624$+$0029   &\nodata \nl

\enddata

\end{deluxetable}

\newpage
\begin{deluxetable}{cllllll}
\tablenum{12}
%\tablewidth{300pt}
\tablecaption{Coefficients of Polynomial Fit to M$_J$:Type and M$_K$:Type}
\tablehead{
\colhead{Magnitude} & \colhead{c$_0$} & \colhead{c$_1$}
& \colhead{c$_2$} & \colhead{c$_3$}
& \colhead{c$_4$} & \colhead{c$_5$} \nl
}
\tablecolumns{7}
\startdata
M$_{J_{MKO}}$ &  12.03 & $-$6.278e$-1$ & 4.500e$-1$ & $-$6.848e$-2$ & 3.986e$-3$ & $-$7.923e$-5$\nl
M$_{K_{MKO}}$ &  10.93 & $-$6.485e$-1$ & 3.876e$-1$ & $-$5.819e$-2$ & 3.524e$-3$ & $-$7.351e$-5$ \nl
\nl
\tablenotetext{}{Fit is valid for L1 to T9 and is applied as 
$$mag  = c_{0} + c_{1}\times Type +
c_{2}\times Type^2  + c_{3}\times Type^3 +
c_{4}\times Type^4  + c_{5}\times Type^5
$$ 
where type is an integer such that 01$=$L1, 10$=$T0, 19$=$T9.}

\enddata

\end{deluxetable}
\end{document}